\documentclass[aps,prd,10pt,showpacs,amsmath,showkeys,twocolumn,floatfix,amssymb, preprintnumbers, nofootinbib, superscriptaddress]{revtex4}
\usepackage{graphicx}
\usepackage{comment}
\usepackage[usenames]{color}
\usepackage{bm}
\usepackage{ifpdf}
\usepackage{floatrow}
\usepackage{makecell}
\usepackage{caption}
\usepackage{stackrel}
\usepackage{subcaption}

\usepackage[normalem]{ulem}
\usepackage[dvipsnames]{xcolor}
\usepackage[utf8]{inputenc}
\usepackage{hyperref}
\hypersetup{
  pdfnewwindow=true,      
  colorlinks=true,        
  linkcolor=PineGreen,    
  citecolor=PineGreen,    
  filecolor=PineGreen,    
  urlcolor=PineGreen      
}
\newcommand{\be}{\begin{eqnarray}}
\newcommand{\ee}{\end{eqnarray}}

\begin{document}

\title{Toward extracting scattering phase shift from integrated correlation function}

\author{Peng~Guo}
\email{peng.guo@dsu.edu}

\affiliation{College of Arts and Sciences,  Dakota State University, Madison, SD 57042, USA}
\affiliation{Department of Physics and Engineering,  California State University, Bakersfield, CA 93311, USA}
\affiliation{Kavli Institute for Theoretical Physics, University of California, Santa Barbara, CA 93106, USA}

\author{Vladimir~Gasparian}
\email{vgasparyan@csub.edu}
\affiliation{Department of Physics and Engineering,  California State University, Bakersfield, CA 93311, USA}

\date{\today}

\begin{abstract}
In present work, a relation that connects the integrated correlation function of a trapped two-particle system   to infinite volume particles scattering phase shift  is derived.  It has the potential to provide an alternative approach for  extracting two-particle scattering phase shift from  integrated correlation function in lattice simulation at small Euclidean time region.
  Both (i)  perturbation calculation of  1+1 dimensional lattice Euclidean field theory model of fermions interacting with a contact  interaction and  (ii) Monte Carlo simulation  of a 1D exactly solvable quantum mechanics model are carried out to test the proposed relation. In contrast to conventional two-step approach of   extracting   energy levels    from temporal correlation function in  lattice simulation at large Euclidean time first and then  applying L\"uscher formula to convert energy levels into scattering phase shifts,  we show that the difference of integrated correlation functions between interacting and noninteracting trapped systems converges rapidly to  infinite volume limit that is given in terms of scattering phase shifts  at small Euclidean time region.
\end{abstract}

\maketitle

\section{Introduction}\label{sec:intro}

Study of  hadron/nuclear particles interactions and properties of few-body resonances from the first principles, quantum Chromodynamics (QCD) that is the underlying theory of quark and gluon interactions,  is one of the major tasks in modern hadron/nuclear  physics. In particular hadron/nuclear particles provide the only means of understanding  the quark and gluon dynamics. However, extracting information of hadron/nuclear particle interactions from the first principles, such as scattering phase shifts, is not always straightforward. Usually theoretical computations are performed  in various traps, for instance,   periodic cubic box in  lattice quantum Chromodynamics (LQCD) and harmonic oscillator trap in nuclear physics.  As the result of trapped systems, the energy spectra become discrete.  To extract the scattering informations,  normally the   two-step procedures are carried out:  (1) First of all, discrete low-lying few-body energy levels are extracted by fitting exponential decaying behavior of correlation functions in Euclidean space-time,  and looking for the plateau  in temporal correlation functions  when Euclidean time is large enough so that all excited energy levels  decay off rapidly and only lowest energy level becomes dominant. The energy spectra of excited states can be extracted in a similar way by applying variational approach and generalized  generalized eigenvalue method \cite{Michael:1985ne,Luscher:1990ck,Blossier:2009kd};  (2)   Applying L\"uscher formula  \cite{Luscher:1990ux}  in LCQD  or  Busch-Englert-Rza\.zewski-Wilkens (BERW) formula \cite{Busch98} in a harmonic oscillator (h.o.) trap in nuclear physics, the discrete energy spectra of trapped system can be converted into scattering phase shifts, etc. The L\"uscher formula and BERW formula have been quickly extended to include inelastic effects, such as coupled-channel effect and three-body problems,  etc.  see e.g. Refs.~\cite{Rummukainen:1995vs,Christ:2005gi,Bernard:2008ax,He:2005ey,Lage:2009zv,Doring:2011vk,Guo:2012hv,Guo:2013vsa,Kreuzer:2008bi,Polejaeva:2012ut,Hansen:2014eka,Mai:2017bge,Mai:2018djl,Doring:2018xxx,Guo:2016fgl,Guo:2017ism,Guo:2017crd,Guo:2018xbv,Mai:2019fba,Guo:2018ibd,Guo:2019hih,Guo:2019ogp,Guo:2020wbl,Guo:2020kph,Guo:2020iep,Guo:2020ikh,Guo:2020spn,Guo:2021lhz,Guo:2021uig,Guo:2021qfu,Guo:2021hrf,Stetcu:2007ms, Stetcu:2010xq,Rotureau:2010uz,Rotureau:2011vf,Luu:2010hw,Yang:2016brl,Johnson:2019sps,Zhang:2019cai, Zhang:2020rhz}.  This two-step approach has been proven very successful in number of applications especially in meson sector, see e.g. Refs.~\cite{Aoki:2007rd,Feng:2010es,Lang:2011mn,Aoki:2011yj,Dudek:2012gj,Dudek:2012xn,Wilson:2014cna,Wilson:2015dqa,Dudek:2016cru,Beane:2007es, Detmold:2008fn, Horz:2019rrn,Guo:2020kph}. However, the two-step approach also display some disadvantages that are summarized nicely in Ref.~\cite{Bulava:2019kbi}, such as, determination of energy levels in large spatial volume becomes difficult, etc. The situation is even more challenging in baryon sector, finding a clear signal of stable plateau in  nucleon-nucleon reaction correlation functions and pulling out energy spectra from the noisy lattice simulation data is already a difficult task.  Therefore, there have been number of proposals to explore alternative approaches in recent years, such as, determining scattering amplitudes from finite-volume spectral functions in  Ref.~\cite{Bulava:2019kbi} and extraction of spectral densities from lattice correlators in Refs.~\cite{Hansen:2019idp,Bailas:2020qmv}, etc.

    In present work,   we will establish a   connection between integrated correlation functions and scattering phase shift  that has the potential to provide an alternative approach of extracting scattering phase shifts from lattice QCD calculation    and show that:

  (i) The difference of  integrated trapped two-particle correlation functions  between interacting particles system and free particles system  in 1+1 space-time dimensions is related to infinite volume particles scattering phase shift, $\delta(\epsilon)$, by
\begin{equation}
  C  (t)  - C_0 (t)     \stackrel[t=-i\tau]{trap \rightarrow \infty}{\rightarrow} \frac{1}{\pi}   \int_0^\infty d \epsilon  \frac{  d \delta (\epsilon) }{d  \epsilon}   e^{-    \epsilon  \tau}  +  \frac{ \delta(0)}{\pi},   \label{mainresult}
\end{equation}
where $C(t)$ and $C_0(t)$ are integrated correlation functions for two interacting and noninteracting particles in a trap respectively, and $\tau$ stands for Euclidean time.

(ii) Integrated  trapped correlation functions are given in terms of eigenenergies of two particles  by
\begin{equation}
  C  (t)  - C_0 (t)     \stackrel{t=-i\tau}{=}  \sum_n [e^{- \epsilon_n \tau} - e^{- \epsilon^{(0)}_n \tau} ],
\end{equation}
where $\epsilon_n$ and $\epsilon_n^{(0)}$ are eigenenergies of two interacting and noninteracting particles in a trap respectively.
Hence  integrated  trapped correlation functions   resemble the partition function  in statistical mechanics, 
\begin{equation}
  C  (t)  - C_0 (t)  \stackrel{ \tau  \leftrightarrow \beta}{  \leftrightarrow } Tr[e^{ - \beta \hat{H} } -e^{ - \beta \hat{H}_0 }  ], \label{statisticanalog}
  \end{equation}
   with $\tau $   playing the role of $\beta = \frac{1}{k_B T}$.  $\hat{H}$ and $\hat{H}_0$ are interacting and noninteracting particles Hamiltonian operators respectively. The relation given in Eq.(\ref{mainresult}) therefore is analogous  to    well-known result in calculation of   second  virial coefficient of quantum gas by virial expansion approach (also known as cluster expansion method) in quantum statistical mechanics, see e.g. Ref.~\cite{Huang_1987,LIU201337}.

(iii) As discussed in Ref.~\cite{LIU201337},  at high temperature, the scattering cross section is of the order  the square of the thermal de Broglie wavelength, which becomes much smaller than average inter-particle distance  in quantum gas systems, hence  inclusion of only few-body correlations in quantum virial expansion has proven already sufficient at   high temperature (small $\beta$) in describing and understanding properties of quantum gas systems.  In a   similar situation,   two distinct physical scales in  integrated  trapped correlation functions are: (1) the Euclidean evolving time $\tau $ that plays the role of square of the thermal de Broglie wavelength  and (2) the size of trap, $L$. When $\tau$ is much smaller than $L$, the difference of integrated trapped two-particle correlation functions  can be described  by series  expansion   in terms of  powers of $\tau/L$,    we may expect that  the difference of integrated trapped two-particle correlation functions rapidly approaches  the infinite volume limit that is  given in terms of scattering phase shifts in Eq.(\ref{mainresult}) at small $\tau $ region even with a modest size of trap.   This conclusion as matter of fact can be easily illustrated by using relation listed in Eq.(\ref{statisticanalog}), near small $\tau \sim 0$, by Taylor expansion  $$e^{- \hat{H} \tau} \sim ( 1 -\hat{V} \tau + \cdots ) e^{- \hat{H}_0 \tau }, $$
where $\hat{V} = \hat{H} - \hat{H}_0$ stands for interaction operator,  we can show that
\begin{equation}
  C  (t)  - C_0 (t)  \stackrel{ \tau \sim 0}{  \sim  }  -  \langle \hat{V} (\tau) \rangle  \tau  + \mathcal{O}  \left ( \langle \hat{V} (\tau) \rangle^2  \tau^2  \right ),   \label{tauoverLrelation}
  \end{equation}
where $\langle \hat{V} (\tau) \rangle = Tr [\hat{V}  e^{- \hat{H}_0 \tau }  ] $ may be interpreted as thermal average of particles interaction that is proportional to the inverse size of trap: $\langle \hat{V} (\tau) \rangle  \propto \frac{1}{L}$. Therefore, as $\tau/L \ll 1$,   thermal de Broglie wavelength is much smaller than size of trap,  particles becomes less aware of finite size   of a trap, and the difference of integrated trapped two-particle correlation functions agree well with the infinite volume limit   result even with  finite size of a trap. This observation is further  illustrated analytically in great details by perturbation calculation of two-fermion correlation function of  a simple lattice field theory model in Sec.~\ref{pertrubationsolution1D}.

(iv) Another one of our primary goals in present work is  thus  firstly to  illustrate numerically  that two sides in  Eq.(\ref{mainresult}) indeed display a rather good agreement at small $\tau$ region even with a modest size of the trap,   as  the size of trap is increased, the agreement then starts expand into larger $\tau$ region, and  secondly to establish the possibility   of extracting    the infinite volume particles scattering phase  from Monte Carlo calculation of integrated  correlation functions of  trapped two-particle system    at small $\tau$ region.     The Monte Carlo simulation test of a quantum mechanical model with a spinless particle interacting with a square well potential in a harmonic trap is carried out in this work in Sec.~\ref{MCsimulation}. Monte Carlo data indeed show a good agreement with infinite volume limit near small $\tau$ region, and the range of agreement in $\tau$ start expanding as the size of trap is increased.

We remark that at current scope, all our discussions are only limited to nonrelativistic dynamics in one spatial and one temporal dimensional space-time, a lot of more studies must to be  conducted to include relativistic dynamics and inelastic effect, etc before it can be applied to realistic cases in lattice QCD calculation.

The paper is organized as follows. First of all,  a field theory model for the study of nonrelativistic fermions interaction in a trap is set up in Sec.~\ref{modelsetup},  the dynamics of  two fermions interaction in a trap and the  two-particle correlation function are also  presented  in Sec.~\ref{modelsetup}.   The derivation of the infinite volume limit of integrated two-particle correlation function, and its relation to particles scattering phase shift are given in  Sec.~\ref{correlationphaseshift}.  The perturbation calculation of two fermions correlation function of a lattice field theory is carried out and  presented in  Sec.~\ref{pertrubationsolution1D}.    The 1+1D Monte Carlo simulation test with a exact solvable quantum mechanics model is presented and discussed in  Sec.~\ref{MCsimulation}.  The discussions and summary are given in Sec.~\ref{summary}.

\section{A field theory model for nonrelativistic  fermions interaction in a trap}\label{modelsetup}

In this section, we first setup a $1+1$ dimensional field theory model for the study of two nonrelativistic fermions interaction in a trap. All the conventions are established in Sec.~\ref{model}, the dynamical equations of two particles interaction are presented in Sec.~\ref{twoparticledynamics}, and the  definition of time forward propagating two-particle correlation function is given in Sec.~\ref{defcorrelationfunc}.

\subsection{A field theory model setup}\label{model}

To restrain our current discussion   in the case of  single species nonrelativistic particles  interaction,  a simple $1+1$  dimensional  nonrelativistic field theory model  of spin-1/2 fermions interaction via a short-range potentail in a trap is adopted in this work.   Hamiltonian operator of the trapped fermions system is  
\begin{align}
\hat{H} &=\sum_{\sigma = \uparrow, \downarrow } \int d x \hat{\psi}^\dag_{\sigma} (x) \left [ - \frac{1}{2m} \frac{d^2}{d x^2}  + U(x)\right ] \hat{\psi}_{\sigma} (x) \nonumber \\
& + \frac{1}{2} \int d x d y  \hat{\psi}^\dag_{\uparrow} (x) \hat{\psi}^\dag_{\downarrow} (y) V(x-y)  \hat{\psi}_{\downarrow} (y) \hat{\psi}_{\uparrow} (x),
\end{align}
where $\sigma= \uparrow, \downarrow $ and $m$ refer to the fermion polarizations and mass respectively, and $ \hat{\psi}_{\sigma} (x) $ stands for the fermion field operator. The trap potential and short-range interaction potential between two fermions with opposite polarizations  are represented by $U(x)$ and $V(x-y)$ respectively.  Only spatially symmetric short-range interaction is considered in this work: $$V(x-y)=V(y-x), $$  hence, the interaction between two fermions with the same polarizations is suppressed by Pauli exclusive principle.

The second quantization representation of Hamiltonian operator can be obtained by using   relations:
\begin{equation}
\hat{\psi}_{\sigma} (x) = \sum_n \varphi_n (x) a_{n, \sigma}, \ \ \ \hat{\psi}^\dag_{\sigma} (x) = \sum_n \varphi^*_n (x) a^\dag_{n, \sigma},
\end{equation}
where $a_{n, \sigma}$ and $a^\dag_{n, \sigma}$ are the annihilation and creation operators for a single fermion state that is labeled by quantum numbers of $(n,\sigma)$. The expansion coefficient, $\varphi_n (x) $,  is the eigen-wave-function of single particle state  in the trap corresponding to the eigenenergy of $\epsilon^{(0)}_n$, and it satisfies Hartree-Fock-like equation, see e.g. Ref.~\cite{Fetter},
\begin{equation}
\left [   - \frac{1}{2m} \frac{d^2}{d x^2}  + U(x) \right ] \varphi_n (x)  = \epsilon^{(0)}_n \varphi_n (x) .
\end{equation}
The second quantization representation of  Hamiltonian of  trapped fermions system is thus given by
\begin{align}
\hat{H} & =\sum_{\sigma  , n }   \epsilon^{(0)}_n  a^\dag_{\sigma, n}  a_{\sigma, n}   \nonumber \\
&+  \frac{1}{2}  \sum_{n_1, n_2, n'_1 , n'_2} V_{n'_1, n'_2; n_1, n_2}  a^\dag_{\uparrow, n_1}  a^\dag_{\downarrow, n_2}  a_{\downarrow, n'_2 }  a_{\uparrow, n'_1} ,
\end{align}
where
 \begin{equation}
 V_{n'_1, n'_2; n_1, n_2}   =  \int d x d y  \varphi^*_{n'_1} (x) \varphi^*_{n'_2} (y)  V(x-y)  \varphi_{n_1}  (x)   \varphi_{n_2} (y)   .
\end{equation}

\subsection{Two fermions interaction  in a trap}\label{twoparticledynamics}
The two-particle state is defined in this subsection and dynamical equation of trapped two interacting fermions system is also presented. We   remark that the subscript of spatial integration of a trapped system, $\int_{trap} d x$,  is suppressed in follows, the spatial integration of a trapped system for a periodic box and harmonic oscillator trap  is understood as 
\begin{equation}
\int  d x  = \begin{cases} \int_{ 0}^{L} d x ,  & \mbox{for a periodic box of size } L,  \\ \int_{- \infty}^{ \infty} d x ,  & \mbox{for a h.o. trap}.   \end{cases}
\end{equation}

\subsubsection{Spin singlet  state of two fermions in a trap}
The state of two  fermions  with a total spin-$S$ and interacting with a short-range potential in a trap is   defined by
\begin{equation}
| E \rangle = \sum_{\sigma_1,\sigma_2}  \int d x_1 d x_2 \Psi_E(x_1, x_2)   \chi^{(S)}_{\sigma_1, \sigma_2} \frac{\hat{\psi}^\dag_{\sigma_1} (x_1) \hat{\psi}^\dag_{\sigma_2} (x_2)}{\sqrt{2}} | 0 \rangle, \label{twopartstate}
\end{equation}
where $\Psi_E(x_1, x_2)  $ and $ \chi^{(S)}_{\sigma_1, \sigma_2}$ are spatial and spin wave functions of two fermions system with total spin-$S$ respectively. The  factor $1/\sqrt{2}$ takes into account the exchange symmetry of two distinguishable fermions.   For  spatially symmetric short-range interaction potentials, such as a contact interaction, the anti-symmetric spatial wave function is highly suppressed. Hence only spin single state with total spin-$(S=0)$     is considered in current work,
\begin{equation}
 \chi^{(S=0)}_{\sigma_1, \sigma_2} = \frac{1}{\sqrt{2}} \left ( \delta_{\sigma_1, \uparrow} \delta_{\sigma_2, \downarrow}-  \delta_{\sigma_1, \downarrow} \delta_{\sigma_2, \uparrow} \right ),
\end{equation}
and spatial wave function is symmetric under exchange of coordinates of two particles,
\begin{equation}
  \Psi_E(x_1, x_2)  =  \Psi_E(x_2, x_1) .  
  \end{equation}
The second quantization representation of two-fermion state is   given by
  \begin{equation}
| E \rangle = \frac{1}{2} \sum_{n_1, n_2} \widetilde{\Psi}_E (n_1, n_2)  \left (  a^\dag_{\uparrow, n_1} a^\dag_{ \downarrow, n_2}- a^\dag_{\downarrow, n_1} a^\dag_{ \uparrow, n_2} \right )  | 0 \rangle,   \label{secondtwopartstate}
\end{equation}
where
\begin{equation}
\widetilde{\Psi}_E (n_1, n_2)  =  \int d x_1 d x_2 \Psi_E(x_1, x_2)   \varphi^*_{n_1} (x_1)  \varphi^*_{n_2}  (x_2)  ,
\end{equation}
and it is symmetric under   exchange of two-particle state indices,
\begin{equation}
\widetilde{\Psi}_E (n_1, n_2)  =  \widetilde{\Psi}_E (n_2, n_1)  . 
\end{equation}
In the basis of  single particle wave functions, the two-fermion spatial wave function is thus given  by
\begin{equation}
\Psi_E(x_1, x_2)   = \sum_{n_1, n_2} \widetilde{\Psi}_E (n_1, n_2)   \varphi_{n_1} (x_1)  \varphi_{n_2}  (x_2)  . \label{waveexpansion}
\end{equation}
The orthogonality of two-fermion states, $$\langle E | E' \rangle = \delta_{E,E'}, $$ yields that the spatial wave functions are orthonormal and
\begin{equation}
  \sum_{n_1, n_2} \widetilde{\Psi}^*_{E'} (n_1, n_2) \widetilde{\Psi}_E (n_1, n_2) = \delta_{E,E'} .
\end{equation}

We remark that our current discussion is restricted  to only two-particle elastic region, so that the Fock space expansion in Eq.(\ref{twopartstate}) and Eq.(\ref{secondtwopartstate}) is only limited to two-particle state contribution, the multi-particle states with number of particles equal or greater than three are all neglected for now.

\subsubsection{Dynamical equation of trapped two-fermion system}

The effective dynamical equation for two-particle state can be derived from the variational principle  by evaluating 
\begin{equation}
\frac{\partial}{\partial \Psi^*_{E'} (x_1,x_2) }   \langle E' | \hat{H}  -E | E \rangle  =0.
\end{equation}
With the help of relation in Eq.(\ref{waveexpansion}), we find that
\begin{equation}
\hat{H}_{eff} \Psi_E(x_1, x_2)  = E \Psi_E(x_1, x_2) ,
\end{equation}
where the effective two-particle Hamiltonian is given by the sum of kinetic terms of   particle in the trap and interaction potential between two particles: 
\begin{equation}
\hat{H}_{eff}  = \hat{H}_{trap}  + V (x_1 -x_2) ,
\end{equation}
where
\begin{equation}
  \hat{H}_{trap}   =  - \frac{1}{2m } \frac{d^2}{d x_1^2} +U(x_1)   - \frac{1}{2m } \frac{d^2}{d x_2^2} +U(x_2) . 
\end{equation}
In the basis of  single particle wave functions, the eigen-energy and eigen-state can be solved by diagonalizing matrix element of effective two-particle Hamiltonian,
\begin{align}
 \left [  \hat{H}_{eff}  \right ]_{n_1, n_2; n'_1, n'_2} =  \delta_{n'_1, n_1} \delta_{n'_2, n_2} (\epsilon^{(0)}_{n_1} + \epsilon^{(0)}_{n_2}) + V_{n_1, n_2; n'_1, n'_2}     .
\end{align}

\subsubsection{ Separation of center of mass and relative motions}
The center of mass (CM) motion can be separated out rather straightforwardly for  some commonly used traps, such as,  periodic box in lattice QCD, harmonic oscillator trap in nuclear physics, etc.  
\begin{equation}
  \hat{H}_{trap}   =  - \frac{1}{2M } \frac{d^2}{d R^2} +U_{CM} ( R)   - \frac{1}{2\mu } \frac{d^2}{d r^2} +U_{rel}(r) ,
\end{equation}
where $$M = 2 m  \ \  \mbox{and} \ \ \mu = \frac{m}{2}$$ are the total mass and reduced mass of two-particle system respectively, and  $$R= \frac{x_1+x_2}{2} \ \  \mbox{and} \ \   r = x_1 -x_2$$ are center of mass and relative coordinates of two particles respectively. The $U_{CM} ( R)$ and $U_{rel}(r)$ represent the trap potentials for center of mass and relative motions respectively.  As a specific example, the harmonic oscillator trap potential for individual particle is  
\begin{equation}
U(x) =  \frac{1}{2} m \omega^2 x^2,
\end{equation}
where $\omega$ is   angular frequency of harmonic oscillator. 
The trap potentials for center of mass and relative motions   have the similar forms but the mass of particle, $m$, must be replaced by total mass and reduced mass respectively,
\begin{equation}
U_{CM}(R) =  \frac{1}{2} M \omega^2 R^2,  \ \ \ \  U_{rel}(r) =  \frac{1}{2} \mu \omega^2 r^2.
\end{equation}

The total wave function is thus the product of center of mass wave function and relative wave function,
\begin{equation}
 \Psi_E(x_1, x_2)  = \psi^{(CM)}_{E_{CM}} (R) \psi_{\epsilon}^{(rel)} (r), \ \ \ \ E= E_{CM} +\epsilon,
\end{equation}
they satisfy Schr\"odinger equations:
\begin{equation}
 \left [ - \frac{1}{2M } \frac{d^2}{d R^2} +U_{CM} ( R) \right ]   \psi^{(CM)}_{E_{CM}} (R) =E_{CM} \psi^{(CM)}_{E_{CM}} (R)   ,
\end{equation}
and
\begin{equation}
 \left [ - \frac{1}{2\mu } \frac{d^2}{d r^2} +U_{rel}(r)  +V(r) \right ] \psi_{\epsilon}^{(rel)} (r)    =\epsilon \psi_{\epsilon}^{(rel)} (r)   .
\end{equation}

\subsection{Two fermions correlation function}\label{defcorrelationfunc}
In lattice QCD, particles interaction are usually studied via evaluating time dependence of  correlation functions numerically from the first principle. To illustrate how the two-particle correlation function is related to particles scattering phase shift,   
  firstly  we define  the   forward time propagating two-particle   correlation function by
\begin{equation}
C(r t; r' 0)|_{t>0} = \theta(t) \langle 0| \mathcal{\widehat{O}}_H (r ,t)  \mathcal{\widehat{O}}^\dag_H (r' ,0)| 0 \rangle.
\end{equation}
The  two-particle creation operator in Heisenberg picture  is given by
\begin{equation}
 \mathcal{\widehat{O}}^\dag_H (r ,t) = e^{i \hat{H} t}  \mathcal{\widehat{O}}^\dag (r)  e^{ -i \hat{H} t} ,
\end{equation}
where  
\begin{equation}
 \mathcal{\widehat{O}}^\dag (r ) =  \int d R   \psi^{(CM)}_{E_{CM}} (R)   \frac{ \hat{\psi}^\dag_{ \uparrow} (x_1) \hat{\psi}^\dag_{ \downarrow } (x_2) -  \hat{\psi}^\dag_{ \downarrow } (x_1) \hat{\psi}^\dag_{\uparrow} (x_2)}{2}   .
\end{equation}
and  CM motion has been projected out in   definition  of $ \mathcal{\widehat{O}}^\dag (r ) $ operator.

Inserting complete energy basis in between two-particle annihilation and creation operators, $$\sum_E | E \rangle \langle E| = 1,$$ and also using Eq.(\ref{twopartstate}), it is straightfoward to show that
\begin{equation}
\langle E |  \mathcal{\widehat{O}}^\dag (r ) | 0 \rangle =\psi_{\epsilon}^{(rel)*} (r) .
\end{equation}
   We thus find 
\begin{equation}
 C(r t; r' 0)|_{t>0} = e^{- i E_{CM} t} C^{(rel)}(r t; r' 0)|_{t>0} ,
\end{equation}
where  the correlation function for the relative motion of two-particle system   is given by
\begin{equation}
 C^{(rel)}(r t; r' 0)|_{t>0} = \theta(t) \sum_{\epsilon} e^{- i   \epsilon t}  \psi_{\epsilon}^{(rel)} (r)\psi_{\epsilon}^{(rel)*} (r'). \label{spectralrepcorrelation}
\end{equation}

Using identity
\begin{equation}
i \int_{- \infty}^\infty \frac{d E }{2\pi} \frac{e^{- i E t}}{E  + i 0}  = \theta (t), \label{integthetat}
\end{equation}
the two-particle correlation function can also be written as
\begin{equation}
 C^{(rel)} (r t; r' 0)|_{t>0} = i \int_{- \infty}^\infty \frac{d  \lambda }{2\pi}   \sum_\epsilon  \frac{  \psi_{\epsilon}^{(rel)} (r)\psi_{\epsilon}^{(rel)*} (r') }{  \lambda - \epsilon + i 0 } e^{- i  \lambda  t} .
\end{equation}
Hence the two-particle correlation function is related to  Green's function of two-particle interaction  in a trap by
\begin{equation}
C^{(rel)}(r t; r' 0)|_{t>0} =i  \int_{- \infty}^\infty \frac{d \lambda}{2\pi}  G^{(trap)}(r,r'; \lambda+ i 0)   e^{- i \lambda  t} , \label{CorrelationtoGreen}
\end{equation}
where the spectral representation of two-particle Green's function is given by
\begin{equation}
G^{(trap)}(r,r'; \lambda)  =     \sum_\epsilon  \frac{  \psi_{\epsilon}^{(rel)} (r)\psi_{\epsilon}^{(rel)*} (r') }{ \lambda  - \epsilon  }  ,
\end{equation}
and it satisfies differential equation,
\begin{equation}
 \left [ \epsilon + \frac{1}{2\mu } \frac{d^2}{d r^2} -U_{rel}(r)  - V(r) \right ] G^{(trap)}(r,r'; \epsilon)   = \delta(r-r').
\end{equation}

\section{Integrated two-particle correlation function and its relation to scattering phase shift}\label{correlationphaseshift}
The detailed derivation of how integrated two-particle correlation function is related to scattering phase shift are presented in Sec.~\ref{derivation}, we show that the infinite volume limit of difference of integrated  correlation functions between two interacting  and non-interacting particles in the trap approaches $$ \frac{1}{\pi}   \int_0^\infty d \epsilon   \frac{  d \delta (\epsilon) }{d  \epsilon}  e^{- i   \epsilon t}  + \frac{\delta(0)}{\pi} ,$$ where $\delta(\epsilon)$ is two-particle scattering phase shift in infinite volume.  The relation is then  illustrated by using a exactly solvable contact interaction   model in both periodic box and harmonic oscillator trap in Sec.~\ref{exactsolution1D}.

\subsection{Integrated two-particle correlation function}
 Using orthogonality of two-particle wave function,    the integrated two-particle correlation function
  is    related to energy spectra simply by
 \begin{equation}
  C^{(rel)}( t)|_{t>0}    =   \int d r  C^{(rel)}(r t; r 0)|_{t>0}   = \theta(t)  \sum_{\epsilon} e^{- i   \epsilon t}  .
\end{equation}

 Using Eq.(\ref{CorrelationtoGreen}), the    integrated two-particle correlation function is therefore also given by
 \begin{equation}
  C^{(rel)}( t)|_{t>0}     =i  \int_{- \infty}^\infty \frac{d \lambda}{2\pi}   Tr[G^{(trap)}( \lambda + i 0) ]  e^{- i  \lambda  t}  , \label{CtImGreen}
\end{equation}
 where the trace of Green's function is defined by
 \begin{equation}
 Tr[G^{(trap)}(  \lambda) ] =  \int d r  G^{(trap)}(r,r; \lambda). 
 \end{equation}

\subsection{Relating integrated correlation function to scattering phase shift}\label{derivation}

\subsubsection{Quantization condition of energy spectra in a trap and infinite volume limit of integrated correlation function}
 
With a   short-range interaction, the quantization condition that determines   discrete energy spectra of the trapped two-particle system can be formulated in a compact form,  for instance,  L\"uscher formula  \cite{Luscher:1990ux} in a periodic cubic box in LCQD  and BERW formula \cite{Busch98} in a harmonic oscillator trap in nuclear physics, 
  \begin{equation}
   \det \left [  \cot \delta  (\epsilon) -  \mathcal{M} (\epsilon ) \right ]=0 \, ,
    \label{eqn:generalQC}
\end{equation}
where  $\delta (\epsilon)$ refers to the diagonal matrix of scattering partial wave phase shifts, and the   matrix function
$ \mathcal{M} (\epsilon ) $  is associated to the geometry and dynamics of trap itself. 
  L\"uscher and BERW  formula    both are the result of      presence of two well separated physical  scales: (1) short-range interaction between two particles and (2) size of trap. Therefore the short-range dynamics that is described by scattering phase shift  and long-range correlation effect due to the  trap can be factorized,   also see recent developments and extension  of  L\"uscher and BERW  formalism beyond two-particle sector and elastic region,  Refs.~\cite{Rummukainen:1995vs,Christ:2005gi,Bernard:2008ax,He:2005ey,Lage:2009zv,Doring:2011vk,Guo:2012hv,Guo:2013vsa,Kreuzer:2008bi,Polejaeva:2012ut,Hansen:2014eka,Mai:2017bge,Mai:2018djl,Doring:2018xxx,Guo:2016fgl,Guo:2017ism,Guo:2017crd,Guo:2018xbv,Mai:2019fba,Guo:2018ibd,Guo:2019hih,Guo:2019ogp,Guo:2020wbl,Guo:2020kph,Guo:2020iep,Guo:2020ikh,Guo:2020spn,Guo:2021lhz,Guo:2021uig,Guo:2021qfu,Guo:2021hrf,Stetcu:2007ms,Stetcu:2010xq, Rotureau:2010uz,Rotureau:2011vf,Luu:2010hw,Yang:2016brl,Johnson:2019sps,Zhang:2019cai, Zhang:2020rhz}.

For a particular partial wave state,  Eq.(\ref{eqn:generalQC}) can be rearranged  to 
\begin{equation}
\delta_l (\epsilon) + \phi_l (\epsilon) = n \pi, \  \ \ \ n \in \mathbb{Z}, \label{QC}
\end{equation} 
where $l$ stands for the angular momentum of system. The expression of   $\phi_l (\epsilon) $ is associated to  matrix elements of $ \mathcal{M} (\epsilon ) $ and may   depends on   other partial wave phase shifts as well. For example, in BERW  formula with a harmonic oscillator trap, see e.g. Ref.~\cite{Busch98,Guo:2021qfu},  the rotational symmetry is well preserved, so that $\left [  \mathcal{M} (\epsilon ) \right ]_{l, l'} = \delta_{l,l'} \mathcal{M}_l (\epsilon ) $, and   $ \phi_l (\epsilon) = - \cot^{-1} \left  [    \mathcal{M}_l (\epsilon ) \right ]$ is totally determined by diagonal element of $\mathcal{M} (\epsilon )$ matrix. However in a periodic cubic box, see e.g. Ref.~\cite{Luscher:1990ux,Guo:2021qfu},  the rotational symmetry is broken and angular orbital momenta are no longer good quantum numbers, hence  $\mathcal{M} (\epsilon )$ in general is not a diagonal matrix. $\phi_l (\epsilon )$ now not only depend on matrix element of $\mathcal{M} (\epsilon )$, but also depends on other partial wave phase shifts as well.

In one   dimensional space, partial wave angular momentum states are replaced by the parity states. In our case, for spin singlet two-fermion states, only even parity state contributes, so from this point on, the subscript-$l$  in Eq.(\ref{QC}) will be dropped, the quantization condition is simply written as
\begin{equation}
\delta (\epsilon_n) + \phi (\epsilon_n) = n \pi,
\end{equation}
 where subscript-$n$ in $\epsilon_n$ is used to label the $n$-th eigen-energy of system. The analytic expression of $ \phi (\epsilon)$ for a periodic box and harmonic oscillator (h.o.) trap are given in Appendix \ref{QC1Dappend} respectively by
\begin{equation}
\phi (\epsilon) = 
 \begin{cases} 
\sqrt{2 \mu \epsilon } \frac{ L}{2} , &   \mbox{for a periodic box} , \\
-\cot^{-1} \left [ \sqrt{ \frac{\epsilon}{2 \omega}}   \frac{\Gamma(\frac{1}{4} - \frac{\epsilon}{2 \omega})}{\Gamma(\frac{3}{4} - \frac{\epsilon}{2 \omega})} \right ] ,&  \mbox{for  a h.o. trap},
 \end{cases} \label{QCphitrap}
\end{equation}
where $L$ stands for the size of periodic box, so that the wave function in CM frame with zero total momentum satisfies periodic boundary condition of
\begin{equation}
\psi_{\epsilon}^{(rel)} (r+n L) = \psi_{\epsilon}^{(rel)} (r), \ \ \ \ n \in \mathbb{Z}. \label{periodicwave}
\end{equation}
The technical details of derivation of Eq.(\ref{QCphitrap}) can be found in Refs.~\cite{Guo:2021uig,Guo:2021lhz,Guo:2020ikh} and also Appendix \ref{QC1Dappend}, the same approach applies to one dimensional case as well.

Using the fact that
 \begin{equation}
\frac{1}{\pi} \left [ \triangle \delta (\epsilon_n) + \triangle \phi (\epsilon_n)  \right ]= 1, \ \ \ \ n=0,1, \cdots,
\end{equation}
where
\begin{equation}
\triangle \delta (\epsilon_n)  = \delta (\epsilon_{n+1}) -   \delta (\epsilon_n) ,  \ \  \triangle \phi (\epsilon_n)  = \phi (\epsilon_{n+1}) -   \phi (\epsilon_n),
\end{equation}
we thus have a relation
\begin{equation}
  \sum_{ n =0}^\infty e^{- i   \epsilon_n t}  = \frac{1}{\pi}  \sum_{ n=0 }^\infty  \triangle \epsilon_n  \left [ \frac{ \triangle \delta (\epsilon_n) }{\triangle \epsilon_n}+ \frac{ \triangle \phi (\epsilon_n) }{\triangle \epsilon_n} \right ]  e^{- i   \epsilon_n t}  ,
\end{equation}
where $ \triangle \epsilon_n =    \epsilon_{n+1} -    \epsilon_n$. When the particle interaction is turned off, $\delta (\epsilon) \rightarrow 0$, the energy spectra is totally determined by condition, $$\phi (\epsilon^{(0)}_n) = n \pi ,$$ where $\epsilon^{(0)}_n$ stands for the $n$-th eigen-energy  of noninteracting two-fermion system in a trap. Hence, we also have a relation, 
\begin{equation}
  \sum_{ n=0 }^\infty e^{- i   \epsilon^{(0)}_n t}  = \frac{1}{\pi}  \sum_{ n=0 }^\infty  \triangle \epsilon^{(0)}_n  \frac{ \triangle \phi (\epsilon^{(0)}_n) }{\triangle \epsilon^{(0)}_n}   e^{- i   \epsilon^{(0)}_n t}  .
\end{equation}

As the system in a trap is approaching infinite volume limit, such as $L \rightarrow \infty$ in a periodic box or $\omega \rightarrow 0$ in harmonic oscillator trap, we find
\begin{equation}
  \sum_{ n =0 }^\infty  e^{- i   \epsilon_n t}  \stackrel{trap \rightarrow \infty}{\rightarrow} \frac{1}{\pi}   \int_0^\infty d \epsilon   \left [ \frac{  d \delta (\epsilon) }{d  \epsilon}+ \frac{ d \phi (\epsilon) }{d \epsilon} \right ]  e^{- i   \epsilon t}  , \label{unsubtractedCtintinflimit}
\end{equation}
and
\begin{equation}
  \sum_{ n=0 }^\infty e^{- i   \epsilon^{(0)}_n t}   \stackrel{trap \rightarrow \infty}{\rightarrow} \frac{1}{\pi}   \int_0^\infty d \epsilon    \frac{ d \phi (\epsilon) }{d \epsilon}    e^{- i   \epsilon t}  .
\end{equation}
Hence   it is tempting to conclude that
\begin{equation}
  \sum_{ n=0 }^\infty  \left [ e^{- i   \epsilon_n t} - e^{- i   \epsilon^{(0)}_n t} \right ] \stackrel{trap \rightarrow \infty}{\rightarrow} \frac{1}{\pi}   \int_0^\infty d \epsilon  \frac{  d \delta (\epsilon) }{d  \epsilon}   e^{- i   \epsilon t}  . \label{diffpartition}
\end{equation}
At the limit of $V(r) \rightarrow 0$,  the left hand side of Eq.(\ref{diffpartition}) clearly approaches zero, on the contrary, on the right hand side of Eq.(\ref{diffpartition}), the weak interaction limit of $ \frac{  d \delta (\epsilon) }{d  \epsilon}  $  is not always well defined, such as in the case of 1D contact interaction potential. However, in general, the weak interaction limit of  $   \delta (\epsilon)   $ is well defined and approaches zero, hence using integration by part, we find
\begin{equation}
   \frac{1}{\pi}   \int_0^\infty d \epsilon  \frac{  d \delta (\epsilon) }{d  \epsilon}   e^{- i   \epsilon t}  \stackrel{V(r) \rightarrow 0}{\rightarrow}  - \frac{\delta(0)}{\pi} ,  
\end{equation}
where $\delta(0)$ is the possible non-trivial surface term as the result of integration by part,   other surface terms are assumed trivial and vanishing.  For instance,  for the particles interacting with a contact interaction in 1D, the phase shift at branch point has non-trivial value: $\delta(0) = - \frac{\pi}{2}$. 
In order to make sure both side of Eq.(\ref{diffpartition})  approach zero at the limit of weak interaction, the   constant shift, $- \frac{\delta(0)}{\pi}$, at  right hand side must be subtracted. 
Therefore, at the infinite volume limit, the difference between integrated correlation functions with and without particle  interactions is associated to scattering phase shift by
\begin{align}
& \left   [ C^{(rel)} (t)  - C^{(rel)}_0 (t) \right ]_{t >0}  \nonumber \\
& \stackrel{trap \rightarrow \infty}{\rightarrow} \frac{\theta(t)}{\pi}   \int_0^\infty d \epsilon  \frac{  d \delta (\epsilon) }{d  \epsilon}   e^{- i    \epsilon  t}  +  \frac{\theta(t) }{\pi} \delta(0), \label{diffCt}
\end{align}
where $C^{(rel)}_0 (t)$ is integrated correlation function of noninteracting particles  in a trap,
\begin{equation}
C^{(rel)}_0 (t) |_{t >0} = \theta(t) \sum_{ n=0 }^\infty e^{- i    \epsilon^{(0)}_n t}  .
\end{equation}
Similarly, the     constant shift, $- \frac{\delta(0)}{\pi}$, must be subtracted in Eq.(\ref{unsubtractedCtintinflimit}) as well,
\begin{equation}
  \sum_{ n =0 }^\infty  e^{- i   \epsilon_n t}  \stackrel{trap \rightarrow \infty}{\rightarrow} \frac{1}{\pi}   \int_0^\infty d \epsilon   \left [ \frac{  d \delta (\epsilon) }{d  \epsilon}+ \frac{ d \phi (\epsilon) }{d \epsilon} \right ]  e^{- i   \epsilon t}  +  \frac{\delta(0) }{\pi} . \label{Ctintinflimit}
\end{equation}

\subsubsection{The role of Friedel formula and Krein's theorem}

The expression in Eq.(\ref{diffCt}) can also be understood by the relation displayed in Eq.(\ref{CtImGreen}).  In terms of Green's function of two-particle system in the trap,  the   difference between integrated correlation functions with and without particle  interactions is also given by
  \begin{align}
& \left   [ C^{(rel)}(t)  - C^{(rel)}_0 (t) \right ]_{t >0}   \nonumber \\
&= i  \int_{- \infty}^\infty \frac{d \lambda}{2\pi}  \left [Tr[G^{(trap)}(  \lambda + i 0) -G_0^{(trap)}(  \lambda + i 0) ] \right ]  e^{- i  \lambda  t}  ,  
\end{align}
where $Tr[G_0^{(trap)}( \lambda+ i0 )]$ stands for the trace of Green's function of two noninteracting particles   in the trap. As the system is approaching infinite volume limit, thus
  \begin{align}
& \left   [ C^{(rel)} (t)  - C^{(rel)}_0 (t) \right ]_{t >0}   \stackrel{trap \rightarrow \infty}{\rightarrow}   \nonumber \\
& i  \int_{- \infty}^\infty \frac{d \lambda}{2\pi}  \left [Tr[G^{(\infty)}(  \lambda + i 0) -G_0^{(\infty)}(  \lambda + i 0) ] \right ]  e^{- i  \lambda  t}   . \label{diffCrelGreen}
\end{align}
As demonstrated in  Refs.~\cite{doi:10.1080/00018735400101233,Friedel1958,zbMATH03313022,krein1953trace,Faulkner_1977} and also see discussion in Ref.~\cite{Guo:2022row}, the difference between the   trace of Green's function  of the interacting system and free particle system is related to the scattering phase shift by Friedel formula and Krein's theorem, see short summary in Appendix \ref{FriedelKreinappend},
 \begin{equation}
 - Tr[G^{(\infty)}( \lambda ) -G_0^{(\infty)}( \lambda ) ]   = \frac{1}{\pi}   \int_0^\infty   d \epsilon    \frac{\delta(\epsilon)}{(\epsilon - \lambda)^2}  , \label{Kreintheorem}
\end{equation}
where we have assumed $Tr[G^{(\infty)}( \lambda )]$ only has a dominant physical branch cut along positive real axis in complex $\lambda$-plane: $\lambda \in [0, \infty]$ and  an unphysical branch cut sitting along negative real axis has been neglected at the scope of current work.   By using Eq.(\ref{Kreintheorem}),  the Eq.(\ref{diffCrelGreen}) thus can be rearranged to
  \begin{align}
& \left   [ C^{(rel)} (t)  - C^{(rel)}_0 (t) \right ]_{t >0}   \nonumber \\
&  \stackrel{trap \rightarrow \infty}{\rightarrow}     -  \frac{1}{\pi}   \int_0^\infty   d \epsilon \delta(\epsilon)   \frac{d}{d \epsilon}  \left [  i  \int_{- \infty}^\infty \frac{d \lambda}{2\pi} \frac{e^{- i  \lambda  t} }{\lambda - \epsilon + i 0 }   \right ]   .
\end{align}
Using identity Eq.(\ref{integthetat}) again, we find
  \begin{align}
  \left   [ C^{(rel)} (t)  - C^{(rel)}_0 (t) \right ]_{t >0}   \stackrel{trap \rightarrow \infty}{\rightarrow}     -  \frac{\theta(t)}{\pi}   \int_0^\infty   d \epsilon \delta(\epsilon)   \frac{d e^{- i  \epsilon  t}  }{d \epsilon}  ,
\end{align}
 hence by integration by part and assuming $\delta(\infty) \rightarrow 0$, Eq.(\ref{diffCt}) is obtained again.

\subsection{1D analytic solutions of two fermions in traps interacting with a contact interaction}\label{exactsolution1D}

  \begin{figure*}
 \centering
 \begin{subfigure}[b]{0.47\textwidth}
\includegraphics[width=0.99\textwidth]{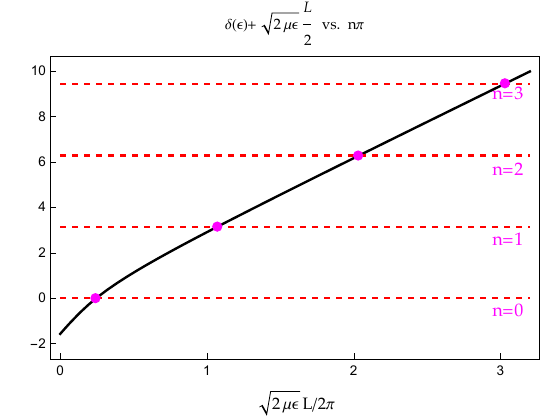}
\caption{  }\label{periodboxplot1}
\end{subfigure} 
\begin{subfigure}[b]{0.49\textwidth}
\includegraphics[width=0.99\textwidth]{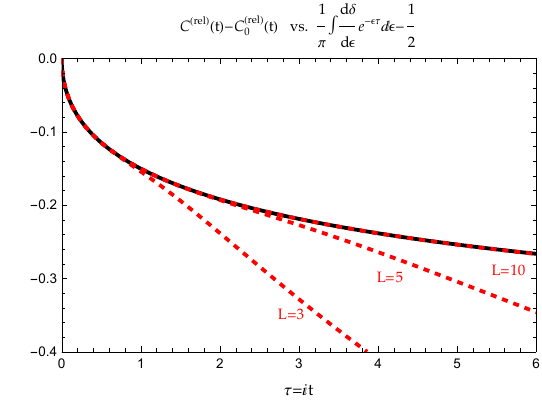}
\caption{    }\label{periodboxplot2}
\end{subfigure}
 \caption{The energy spectra and difference of integrated correlation function plots for particles interaction in a periodic box: (a) $\delta(\epsilon_n) + \sqrt{2\mu \epsilon_n} \frac{L}{2} $ (solid black) vs. $n \pi$ (dashed red) with $L=3$, energy spectra are located at intersection points of black and red curves; (b) $   \frac{1}{\pi}  \int_0^\infty d \epsilon  \frac{d \delta(\epsilon)}{d \epsilon}     e^{-  \epsilon    \tau}  -\frac{1}{2}$  (solid black) vs. $C^{(rel)} (t) -C_0^{(rel)} (t) $ (dashed red)  with $L=3,5,10$. The rest of parameters are taken as:       $V_0=0.5$ and $\mu=1$.  \label{periodicboxfigs} }
 \end{figure*}

In this subsection, the infinite volume limit of difference of integrated correlation functions between two interacting and non-interacting particles in the trap is illustrated by considering a simple contact interaction model. The scattering phase shift of particles interaction in infinite volume is exactly solvable. Both a periodic box   and a harmonic oscillator trap are considered,  the derivation of  analytic expression of quantization conditions in both cases are presented in Appendix \ref{QC1Dappend}. The discrete energy spectra of particles interaction in a trap can be solved rather straightforwardly numerically. We will show that the difference of integrated correlation functions in both cases approaches the same infinite volume limit that is solely determined by particles interaction  steadily and converge rather fast  near small $t$ region.

\subsubsection{Periodic box}
Let's first consider a simple problem of 1D two-fermion of total spin-zero interacting with a contact interaction in a periodic box, the dynamics of two-particle system in CM frame  is described by
\begin{equation}
\left [ - \frac{1}{2\mu} \frac{d^2}{d r^2} + V_0 \sum_{n \in \mathbb{Z}} \delta (r+ nL) \right] \psi^{(rel)}_\epsilon (r) =\epsilon \psi^{(rel)}_\epsilon (r), \label{schrodingerbox}
\end{equation}  
where $V_0$ is the strength of contact interaction and $L$ is the size of periodic box. The wave function is symmetric under spatial inversion, $\psi^{(rel)}_\epsilon (-r)=\psi^{(rel)}_\epsilon (r)$, and also  must satisfy periodic boundary condition in Eq.(\ref{periodicwave}). For a contact interaction, the analytic expression of QC can be obtained, see e.g. Refs.~\cite{Guo:2016fgl,Guo:2017ism,Guo:2017crd,Guo:2018xbv},
\begin{equation}
\delta(\epsilon_n) + \sqrt{2\mu \epsilon_n} \frac{L}{2} = n \pi, \ \ \ \  n  =0, 1, \cdots, \label{QCperiodbox}
\end{equation}
where the analytic expression of phase shift is 
\begin{equation}
\delta(\epsilon)  = \cot^{-1} \left ( - \frac{\sqrt{2\mu \epsilon}}{\mu V_0} \right ). \label{deltaphase}
\end{equation}

The integrated correlation function of trapped system in Euclidean time, $t = - i \tau$, is defined by
 \begin{equation}
    C^{(rel)} (t) -C_0^{(rel)} (t)   \stackrel{t=-i \tau }{=}    \sum_{n  = 0}^\infty  \left [  e^{- \epsilon_n \tau}  -    e^{- \epsilon^{(0)}_n \tau} \right ]  ,
\end{equation}
where the energy spectra of interacting trapped system, $\epsilon_n$, are determined by Eq.(\ref{QCperiodbox}), see e.g. Fig.~\ref{periodboxplot1}.     $$\epsilon^{(0)}_n = \frac{1}{2\mu} (\frac{2\pi n}{L})^2$$  is energy spectrum of non-interacting   particles in a periodic box. $C_0^{(rel)} (t) $ can be evaluated analytically,
 \begin{equation}
     C_0^{(rel)} (t)      \stackrel{t=-i \tau }{=} \frac{1}{2}+  \frac{1}{2}  \vartheta_3 ( e^{ - ( \frac{2\pi}{L} )^2 \frac{\tau}{2\mu} })  \stackrel{L \rightarrow \infty}{\rightarrow}      \sqrt{ \frac{\mu}{2\pi \tau}  }  \frac{L}{2},
\end{equation}
where $ \vartheta_3 (z)$ is Jacobi elliptic theta function \cite{NIST:DLMF}. At the limit of large volume, according to Eq.(\ref{Ctintinflimit}), we also have
 \begin{equation}
    \sum_{n  = 0}^\infty e^{- \epsilon_n \tau}  \stackrel[t=-i \tau ]{L \rightarrow \infty}{\rightarrow}      \frac{1}{\pi}  \int_{0}^\infty d \epsilon \left [  \frac{d \delta (\epsilon)}{d \epsilon}   + \frac{\mu}{\sqrt{2\mu \epsilon}} \frac{L}{2}  \right ]   e^{-    \epsilon \tau}  + \frac{\delta(0)}{\pi} .
\end{equation}
Using Eq.(\ref{deltaphase}), the analytic expression on the right hand side of above equation can be obtained
 \begin{equation}
    C^{(rel)} (t)  \stackrel[t=-i \tau ]{L \rightarrow \infty}{\rightarrow}   \frac{1}{2}  \mbox{erfc} (\mu V_0 \sqrt{\frac{\tau}{2\mu}}) e^{(\mu V_0)^2 \frac{\tau}{2\mu}} + \sqrt{ \frac{\mu}{2\pi \tau}  } \frac{L}{2} - \frac{1}{2}  .
\end{equation}
We can now see clearly that both  $C_0^{(rel)} (t) $ and $C^{(rel)} (t) $ have the same divergent  behavior,   $\sqrt{ \frac{\mu}{2\pi \tau}  } \frac{L}{2}$,  at the limit of    $L \rightarrow \infty$ and   also $\tau \rightarrow 0$.
However, the divergence cancel out completely between them, hence we find
 \begin{equation}
     C^{(rel)} (t)  - C_0^{(rel)} (t)   \stackrel[t=-i \tau ]{L \rightarrow \infty}{\rightarrow}   \frac{1}{2}  \mbox{erfc} (\mu V_0 \sqrt{\frac{\tau}{2\mu}}) e^{(\mu V_0)^2 \frac{\tau}{2\mu}}  - \frac{1}{2}  , \label{analyresultperiodicbox}
\end{equation}
see Fig.~\ref{periodboxplot2} for the comparison of difference of integrated correlation functions in a periodic box vs. infinite volume limit.

  \begin{figure*}
 \centering
 \begin{subfigure}[b]{0.47\textwidth}
\includegraphics[width=0.99\textwidth]{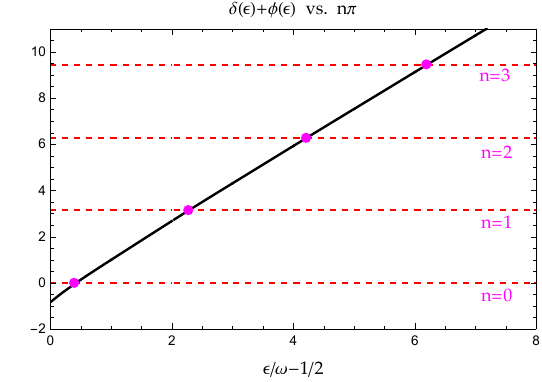}
\caption{  }\label{hotrapplot1}
\end{subfigure} 
\begin{subfigure}[b]{0.49\textwidth}
\includegraphics[width=0.99\textwidth]{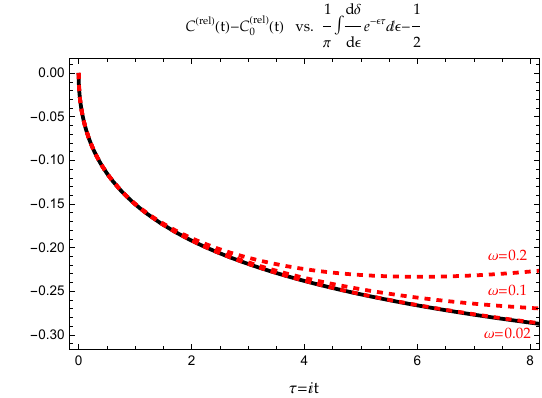}
\caption{    }\label{hotrapplot2}
\end{subfigure}
 \caption{The energy spectra and difference of integrated correlation function plots for particles interaction in a harmonic oscillator trap: (a) $\delta(\epsilon_n) +\phi(\epsilon_n)$ (solid black) vs. $n \pi$ (dashed red) with $\omega=0.2$, energy spectra are located at intersection points of black and red curves; (b) $   \frac{1}{\pi}  \int_0^\infty d \epsilon  \frac{d \delta(\epsilon)}{d \epsilon}     e^{-  \epsilon    \tau}  -\frac{1}{2}$  (solid black) vs. $C^{(rel)} (t) -C_0^{(rel)} (t) $ (dashed red)  with $\omega=0.02,0.1,0.2$. The rest of parameters are taken as:       $V_0=0.5$ and $\mu=1$.  \label{hotrapfigs} }
 \end{figure*}

\subsubsection{Harmonic oscillator trap}

Next let's   consider two-fermion of total spin-zero interacting with a contact interaction in a harmonic oscillator trap, the dynamics of two-particle system    is described by
\begin{equation}
\left [ - \frac{1}{2\mu} \frac{d^2}{d r^2}  + \frac{1}{2} \mu \omega^2 r^2 + V_0   \delta (r) \right] \psi^{(rel)}_\epsilon (r) =\epsilon \psi^{(rel)}_\epsilon (r). \label{schrodingerhotrap}
\end{equation}  
  The spatial wave function again must be symmetric under spatial inversion, $\psi^{(rel)}_\epsilon (-r)=\psi^{(rel)}_\epsilon (r)$, then only even-parity solutions will contribute to the difference of integrated correlation functions. The analytic expression of QC is still given by, see e.g. Refs.~\cite{Guo:2021lhz,Guo:2021uig},
$$ \delta(\epsilon_n)  + \phi(\epsilon_n) = n \pi, $$
where   $ \phi(\epsilon) $ is assumed a smooth monotonically varying function,
\begin{equation}
 \phi(\epsilon)  = -\cot^{-1} \left [ \sqrt{ \frac{\epsilon}{2 \omega}}   \frac{\Gamma(\frac{1}{4} - \frac{\epsilon}{2 \omega})}{\Gamma(\frac{3}{4} - \frac{\epsilon}{2 \omega})} \right ]  + l \pi, \ \ \ l \in \mathbb{Z} . \label{phihotrap}
\end{equation}
The $l \pi$ is added to  keep $\phi(\epsilon) $ monotonically when  $\cot^{-1} (z)$  starts jumping between branches,  see e.g. Fig.~\ref{hotrapplot1}.   Asymptotically $ \phi(\epsilon) $ thus behaves as
\begin{equation}
 \phi(\epsilon)  \stackrel{ \omega \rightarrow 0}{\rightarrow}   \begin{cases} 
 \pi ( \frac{\epsilon}{2\omega} - \frac{1}{4}) , &    \epsilon \gg \omega, \\
 - \frac{\pi}{2} + \frac{\Gamma[\frac{1}{4}]}{\Gamma[\frac{3}{4}]} \sqrt{ \frac{\epsilon}{2\omega}  }  ,& \epsilon  \ll \omega.
 \end{cases} \label{phiasympform}
\end{equation}

The  non-interacting energy spectra of h.o. trap are $$\epsilon^{(0)}_{n} = \omega( n+ \frac{1}{2} ) , $$ hence for even parity solutions, we have
\begin{equation}
  \sum_{n =0 }^{\infty} e^{ - \omega( 2n+ \frac{1}{2} ) \tau} = \frac{1}{2}  \text{csch}  ( \omega \tau ) e^{\frac{\omega \tau}{2}}   \stackrel{\omega \rightarrow  0}{\rightarrow}  \frac{1}{2 \omega \tau}  .
\end{equation}
Using asymptotic form of $\phi(\epsilon) $  in Eq.(\ref{phiasympform}), we find
 \begin{align}
&  \frac{1}{\pi}  \int_{0}^\infty d \epsilon    \frac{d \phi (\epsilon)}{d \epsilon}      e^{-    \epsilon \tau}    \nonumber \\
&  \simeq   \frac{1}{\pi}  \int_{0}^{\omega}  d \epsilon    \frac{d \phi (\epsilon)}{d \epsilon}      e^{-    \epsilon \tau}  + \frac{e^{- \omega \tau}}{2\omega \tau}  \stackrel{\omega \rightarrow  0}{\rightarrow}  \frac{1}{2 \omega \tau}   ,
\end{align}
and
 \begin{equation}
  \sum_{n =0 }^{\infty} e^{ -\epsilon_{n} \tau}  \stackrel{\omega \rightarrow 0}{\rightarrow}   \frac{1}{2}  \mbox{erfc} (\mu V_0 \sqrt{\frac{\tau}{2\mu}}) e^{(\mu V_0)^2 \frac{\tau}{2\mu}} +  \frac{1}{2 \omega \tau}- \frac{1}{2}  .
\end{equation}
Similarly in harmonic oscillator trap,   both  $C_0^{(rel)} (t) $ and $C^{(rel)} (t) $ again show the exact same divergence,  $\frac{1}{2 \omega \tau}$,  at the limit of    $\omega \rightarrow 0$ and   also $\tau \rightarrow 0$.
Hence   after cancellation of divergence,   we find again
 \begin{equation}
     C^{(rel)} (t)  - C_0^{(rel)} (t)   \stackrel[t=-i \tau ]{\omega \rightarrow  0}{\rightarrow}   \frac{1}{2}  \mbox{erfc} (\mu V_0 \sqrt{\frac{\tau}{2\mu}}) e^{(\mu V_0)^2 \frac{\tau}{2\mu}}  - \frac{1}{2}  ,
\end{equation}
see Fig.~\ref{hotrapplot2} for the comparison of difference of integrated correlation functions in the harmonic oscillator trap vs. infinite volume limit.

\subsection{A short summary}
Now we can see clearly that regardless the type of traps that are used, at infinite volume limit where the size of trap is much larger than the range of interaction, the  difference of integrated correlation functions of trapped systems all approach the same limit,
 \begin{equation}
     C^{(rel)} (t)  - C_0^{(rel)} (t)    \stackrel[t=-i \tau ]{trap \rightarrow  \infty }{\rightarrow}  \frac{1}{\pi}   \int_0^\infty d \epsilon   \frac{  d \delta (\epsilon) }{d  \epsilon}  e^{-    \epsilon \tau}  + \frac{\delta(0)}{\pi} , \label{dCtcontinuumlimit}
\end{equation}
   where  the phase shift is given by $$\delta(\epsilon)  = \cot^{-1} \left ( - \frac{\sqrt{2\mu \epsilon}}{\mu V_0} \right )$$ for contact interaction potential. The   analytic expression of infinite volume limit is given by
 \begin{equation}
    \frac{1}{\pi}   \int_0^\infty d \epsilon   \frac{  d \delta (\epsilon) }{d  \epsilon}  e^{-    \epsilon \tau}  + \frac{\delta(0)}{\pi} 
      =   \frac{1}{2}  \mbox{erfc} (\mu V_0 \sqrt{\frac{\tau}{2\mu}}) e^{(\mu V_0)^2 \frac{\tau}{2\mu}}  - \frac{1}{2}  . \label{analyticcontinuumlimit}
\end{equation}
As also illustrated in periodic box and harmonic oscillator trap examples, though both $ C^{(rel)} (t)  $ and $C_0^{(rel)} (t)    $ are divergent as $\tau \sim 0$, the divergence is cancelled out exactly in the difference of two, hence  $ C^{(rel)} (t)   - C_0^{(rel)} (t)   \stackrel{ \tau  \sim  0 }{\rightarrow}  0  $ smoothly.

\section{Perturbation calculation of two fermions correlation function of a lattice field theory model}\label{pertrubationsolution1D}

In this section,  we consider a lattice field theory model of fermions interacting with a contact interaction in a periodic box. For the weak interaction coupling strength, the perturbation calculation of    two fermions correlation function can be carried out directly in path integral representation.  We demonstrated that the infinite volume limit of difference of integrated correlation functions indeed approaches analytic result in Eq.(\ref{dCtcontinuumlimit}) and Eq.(\ref{analyticcontinuumlimit}).

The two fermions correlation function in lattice theory usually is computed in Euclidean space-time by path integral representation,
\begin{align}
& C^{(rel)}(r, t ; r', 0)   =  \frac{ \int \mathcal{D} \psi \mathcal{D} \psi^\dag \hat{O} (r, \tau) \hat{O}^\dag (r', 0) e^{-S_E[\psi, \psi^\dag]} }{\int \mathcal{D} \psi \mathcal{D} \psi^\dag  e^{-S_E[\psi, \psi^\dag]} },
\end{align}
where again $t = - i \tau$.    The relative motion of two-particle creation operator is projected by
\begin{align}
& \hat{O}^\dag (r , \tau)  \nonumber \\
&= \int_{0}^L \frac{d x_2}{\sqrt{L}} \frac{  \psi^\dag_\uparrow (r+ x_2,\tau)    \psi^\dag_\downarrow ( x_2,\tau) - \psi^\dag_\downarrow (r+ x_2,\tau)    \psi^\dag_\uparrow ( x_2,\tau)  }{2} .
\end{align}
The Euclidean action for fermions interacting with a contact interaction in a periodic box with size of $L$ is defined by
\begin{align}
& S_E   [\psi, \psi^\dag] = S_0  + S_V  , \nonumber \\
& S_0   =\int_{- \infty}^\infty  d \tau \int_0^L d x   \sum_{\sigma = \uparrow, \downarrow}   \psi_{\sigma}^\dag (x,\tau) \left (   \partial_\tau   -     \frac{\nabla^2}{2 m }      \right )\psi_{\sigma } (x,\tau) ,  \nonumber \\
 & S_V    = \frac{V_0}{2} \int_{- \infty}^\infty  d \tau \int_0^L d x      \psi^\dag_\uparrow ( x,\tau)   \psi_\uparrow ( x,\tau)  \psi^\dag_\downarrow ( x,\tau)     \psi_\downarrow( x,\tau)    . \label{Euclideanaction}
\end{align}
The fermion field operators satisfy periodic boundary condition,
\begin{equation}
 \psi_\sigma (x+L,\tau)   =  \psi_\sigma (x,\tau) , \ \ \ \   \psi^\dag_\sigma (x+L,\tau)   =  \psi^\dag_\sigma (x,\tau) . 
\end{equation}
We remark that in current scope of discussion, zero lattice spacings in both spatial and temporal directions are assumed.    The size of lattice extent in temporal direction is  also considered infinitely large. Hence lattice artifacts, such as finite lattice spacings  and thermal effect in finite lattice size in temporal direction, etc., are avoided in current discussion. The focus of current work is thus given to  the finite volume effect of correlation function of  a trapped fermions system in a periodic box and its infinite volume limit.

The complete and analytic solutions of two fermions correlation function in path integral representation in field theory seems like a formidable task in general. Fortunately for the weak interaction,  $V_0 \sim 0$, perturbation theory can be carried out. The leading order effect of two fermions correlation function can be obtained rather straightforwardly, and the higher order effects can be carried out systematically  in principle. In current work, only leading order effect of two fermions correlation function is evaluated, and we show that its infinite volume limit indeed is consistent with perturbation expansion of Eq.(\ref{analyticcontinuumlimit}).

 \begin{figure}
\begin{center}
\includegraphics[width=0.99\textwidth]{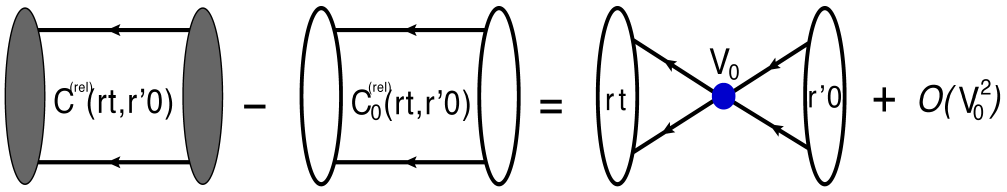}
\caption{   Diagrammatic representation of perturbation calculation  in   Eq.(\ref{perturbationcorrelation}).   }\label{pertbdiagram}
\end{center}
\end{figure}

\subsection{The leading order effect of perturbation calculation}
For weak interaction, $V_0 \sim 0$, two-particle correlation function can be computed analytically by perturbation expansion of $$e^{-S_E} \sim (1- S_V + \cdots) e^{-S_0} .$$ The leading order result is thus given by
\begin{align}
& C^{(rel)}( r, t; r', 0 )  -  C_0^{(rel)}( r, t; r' ,0 )  \nonumber \\
&  = - \frac{   \int \mathcal{D} \psi \mathcal{D} \psi^\dag    \hat{O} (r, \tau) \hat{O}^\dag (r', 0) S_V   e^{-S_0 }  }{  \int \mathcal{D} \psi \mathcal{D} \psi^\dag   e^{-S_0 }}  + \mathcal{O} (V_0^2).
\end{align}
  Working out all the Wick contractions, we thus find
\begin{align}
& C^{(rel)}( r, t; r', 0 )  -  C_0^{(rel)}( r, t; r' ,0 )  \nonumber \\
&  = - \frac{V_0}{L} \int_{0}^L  d x_2  \int_{0}^L  d x'_2  \int_{-\infty}^\infty d \tau'' \int_0^L d x''   \nonumber \\
& \times    D_0^{-1} (r+ x_2 -x'' , \tau - \tau''  ) D^{-1}_0 (x_2 -x'' , \tau -\tau'' )   \nonumber \\
& \times    D_0^{-1} (x'' - r'-  x'_2 ,  \tau''  ) D^{-1}_0 (x'' -x'_2 ,   \tau'' )+ \mathcal{O} (V_0^2) , \label{perturbationcorrelation}
\end{align}
where  free single fermion propagator is defined by 
\begin{align}
& \delta_{\sigma, \sigma'} D_0^{-1} (x -x' , \tau -  \tau' )    \nonumber \\
& =   \frac{   \int \mathcal{D} \psi \mathcal{D} \psi^\dag    \psi_\sigma (x, \tau)  \psi_{\sigma'}^\dag(x', \tau')   e^{-S_0 }  }{  \int \mathcal{D} \psi \mathcal{D} \psi^\dag   e^{-S_0 }}  . \label{invD0expression}
\end{align}
The diagrammatic representation of Eq.(\ref{perturbationcorrelation}) is illustrated  in Fig.~\ref{pertbdiagram}.

Using $S_0$ in Eq.(\ref{Euclideanaction}), and Fourier expansion of free fermion field operator in Euclidean space-time,
\begin{equation}
\psi_\sigma (x,\tau) = \int_{-\infty}^\infty  \frac{d \omega }{2\pi} \frac{1}{L} \sum_{k = \frac{2\pi n}{L}, n \in \mathbb{Z}} e^{i \omega \tau} e^{i k x} \widetilde{ \psi}_\sigma (k,\omega),
\end{equation}
  free single fermion propagator can be worked out rather straightforwardly, we   find 
\begin{align}
& D_0^{-1} (x -x' , \tau -  \tau' )    \nonumber \\
&= \int_{-\infty}^\infty  \frac{d \omega }{2\pi} \frac{1}{L} \sum_{k = \frac{2\pi n}{L}, n \in \mathbb{Z}}  \frac{ e^{i \omega ( \tau - \tau' ) } e^{i k (x-x') } }{i \omega + \frac{k^2}{2m}} .
\end{align}
Therefore we also find
\begin{align}
& C^{(rel)}( r, t; r', 0 )  -  C_0^{(rel)}( r, t; r' ,0 )  \nonumber \\ 
&= - V_0      \int_{-\infty}^\infty  \frac{d \omega }{2\pi}   e^{i  \omega  \tau   }   G^{(L)}_0 (r; - i \omega)  G^{(L)}_0 (r' ; - i \omega) . \label{twocorrpertb}
\end{align}
where two-fermion Green's function in a periodic box is defined by
\begin{equation}
G^{(L)}_0 (r; - i \omega) =  \frac{1}{L} \sum_{k = \frac{2\pi n}{L}, n \in \mathbb{Z}}   \frac{ e^{i k r  } }{ - i  \omega  - \frac{k^2}{2\mu}},
\end{equation}
the analytic expression of $G^{(L)}_0  $ function is given by Eq.(\ref{G0Lperiodicbox}).

 \begin{figure}
\begin{center}
\includegraphics[width=0.99\textwidth]{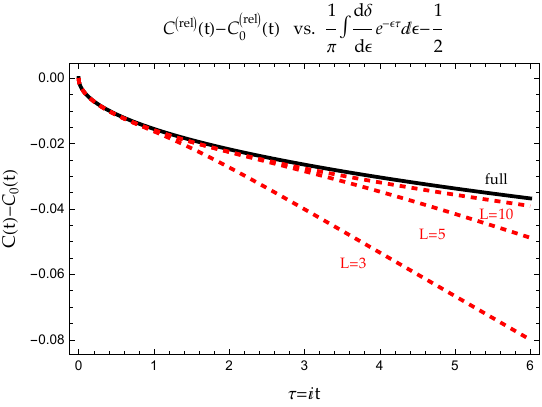}
\caption{  Perturbation calculation of $C^{(rel)} (t) -C_0^{(rel)} (t) $ (dashed red) with $L=3,5,10$ vs. full result in infinite volume limit  $   \frac{1}{\pi}  \int_0^\infty d \epsilon  \frac{d \delta(\epsilon)}{d \epsilon}     e^{-  \epsilon    \tau}  -\frac{1}{2}$ (solid black). The rest of parameters are taken as:       $V_0=0.04$ and $\mu=1$.      }\label{pertbplot}
\end{center}
\end{figure}

\subsection{Integrated two-fermion correlation function and its infinite volume limit}

With perturbation result in Eq.(\ref{twocorrpertb}),  the difference of integrated two-fermion correlation function
\begin{align}
& C^{(rel)}(  t  )    -C_0^{(rel)}(  t  )    \nonumber \\
& = \int_0^L \left [ C^{(rel)}( r, t; r,0  )    -C_0^{(rel)}( r ,t;r, 0  )   \right ]
\end{align}
is thus given by
\begin{align}
 &   C^{(rel)}(  t  )    -C_0^{(rel)}(  t  )     \nonumber \\
 &=  V_0    \sum_{k = \frac{2\pi n}{L}, n \in \mathbb{Z}}  \int_{-\infty}^\infty  \frac{d \omega }{2\pi}   e^{i  \omega  \tau   }    \frac{d}{d (i \omega)} \left   (  \frac{ 1}{i  \omega  + \frac{k^2}{2\mu}}  \right  )   ,
\end{align}
the integration of $\omega$ can be carried out by integration by part, we thus find
\begin{equation}
    C^{(rel)}(  t  )    -C_0^{(rel)}(  t  )      = - \tau  \frac{ V_0 }{L}   \vartheta_3 ( e^{ - (\frac{2\pi}{L})^2 \frac{ \tau}{2\mu} }  )  , \label{ctrelperturbation}
\end{equation}
where $$  \vartheta_3 ( e^{ - (\frac{2\pi}{L})^2 \frac{ \tau }{2\mu} }  )  =   \sum_{k = \frac{2\pi n}{L}, n \in \mathbb{Z}}     e^{  - \frac{k^2}{2\mu}  \tau   }  $$ is Jacobi elliptic theta function \cite{NIST:DLMF}. As $L \rightarrow \infty$, perturbation calculation indeed approach 
\begin{equation}
   C^{(rel)}(  t  )    -C_0^{(rel)}(  t  )   \stackrel[t=-i \tau ]{ L \rightarrow \infty}{  \rightarrow } -  \tau V_0   \int_{- \infty}^\infty  \frac{d k}{2\pi}          e^{-  \frac{k^2}{2\mu} \tau}  = -    V_0  \frac{\sqrt{\mu  \tau}}{\sqrt{2\pi   }}  .
\end{equation}
This is indeed consistent with the perturbation expansion of analytic result in infinite volume limit in  Eq.(\ref{analyticcontinuumlimit}),
 \begin{equation}
   \frac{1}{2}  \mbox{erfc} (\mu V_0 \sqrt{\frac{\tau}{2\mu}}) e^{(\mu V_0)^2 \frac{\tau}{2\mu}}  - \frac{1}{2}    \stackrel{V_0 \rightarrow  0}{\rightarrow}  - V_0 \frac{\sqrt{\mu \tau }}{\sqrt{2\pi}} + \mathcal{O} (V_0^2) ,
\end{equation}
also see     Fig.~\ref{pertbplot} for an example of perturbation calculation result vs. the full result in infinite volume limit.

\subsection{Leading order contribution of energy levels of two-fermion system in finite volume}
As a separate check, we can also evaluate leading order contribution of  energy levels of two-fermion system in perturbation theory in finite volume. The individual energy level can be projected out by
\begin{align}
&  C^{(rel)} (\tau, \epsilon_n ) \nonumber \\
&=  \frac{\sigma_n}{L} \int_{0}^L d r    \int_{0}^L   d r'   e^{ - i k_n (r-r' )}   C^{(rel)}( r, t; r' ,0 )       , \label{Cnprojection}
\end{align}
where $k_n = \frac{2\pi n}{L}, n \in \mathcal{Z}$ is free particle's momentum in finite volume, and $\epsilon_n = \frac{k_n^2}{2\mu} +   \delta  \epsilon_n $ is perturbation result of total energy of two fermions system, where $ \delta \epsilon_n$ stands for the energy shift due to interaction.  The  $\sigma_n $ is degeneracy factor
\begin{equation}
  \sigma_n = \begin{cases}    2      , &  if \ \  n>0 \\     1 , &  if \ \  n=0  . \end{cases}
\end{equation}
 since both $k_{\pm n} = \pm k_n $ correspond to the same free two-fermion energy level:  $\epsilon^{(0)}_n = \frac{k_n^2}{2\mu}$.
 The leading order result of energy shift can also be obtained by using quantization condition in  Eq.(\ref{QCperiodbox}) and Eq.(\ref{deltaphase}),  up to  order of $V_0$, we thus find
\begin{equation}
\delta  \epsilon_n  = \sigma_n \frac{V_0}{L}  .
\end{equation}
 The same conclusion can be obtained by    considering projecting out energy levels from two-fermion correlation function.

First of all, using Eq.(\ref{twocorrpertb}) and Eq.(\ref{Cnprojection}),  we thus find
\begin{equation}
 C^{(rel)} (\tau, \epsilon_n ) -  C_0^{(rel)} (\tau, \epsilon_n ) =    - \tau  \frac{ \sigma_n V_0}{L}   e^{ -   \frac{k_n^2}{2\mu}  \tau   }  +   \mathcal{O} ( \frac{V^2_0}{L^2}  )      ,
\end{equation}
and it is related to Eq.(\ref{ctrelperturbation}) by
\begin{equation}
C^{(rel)}(  t  )    -C_0^{(rel)}(  t  )      =  \sum_{n=0}^\infty \left [ C^{(rel)} (\tau, \epsilon_n ) -  C_0^{(rel)} (\tau, \epsilon_n ) \right ] .
\end{equation}

Next, we also need to consider  the free two-fermion correlation function that is defined by
\begin{equation}
    C_0^{(rel)}( r, t; r' ,0 )    =  \frac{   \int \mathcal{D} \psi \mathcal{D} \psi^\dag  \hat{O} (r, \tau) \hat{O}^\dag (r', 0)     e^{-S_0 } }{  \int \mathcal{D} \psi \mathcal{D} \psi^\dag   e^{-S_0 }}       ,
\end{equation}
and  is given explicitly  in terms of free fermion propagators by
\begin{align}
&    C_0^{(rel)}( r, t; r' ,0 )   =  \frac{1}{2 L} \int_{0}^L  d x_2   \int_{0}^L  d x'_2      \nonumber \\
&   \times     \bigg [  D_0^{-1} ( r + x_2   - r'-  x'_2  , \tau  ) D^{-1}_0 (x_2 - x'_2  , \tau )   \nonumber \\
&  +     D_0^{-1} ( r + x_2  -  x'_2 ,  \tau   ) D^{-1}_0 (x_2  - r'- x'_2 ,   \tau) \bigg ]  .
\end{align}
Using Eq.(\ref{invD0expression}), we thus find
 \begin{equation}
    C_0^{(rel)}( r, t; r' ,0 )    = \frac{1}{L}  \sum_{ k_n= \frac{2\pi n}{L}, n \in \mathbb{Z}} \cos ( k_n r ) \cos (k_n r' )  e^{ -   \frac{k_n^2}{2\mu}  \tau   }   .
\end{equation}
This is indeed consistent with spectral representation of correlation function in Eq.(\ref{spectralrepcorrelation}), where the relative wave function of two free particles in a periodic box is given by 
\begin{equation}
\psi^{(L)}_{k_n} (r) = \frac{1}{\sqrt{L}} \cos (k_n r) ,
\end{equation}
which is normalized by
\begin{equation}
 \int_{0}^L d r  \psi^{(L)}_{k_n} (r) \psi^{(L)*}_{k'_n} (r) = \frac{\delta_{k_n, k'_n} + \delta_{k_n, -k'_n}}{2} = \frac{\delta_{|k_n|, |k'_n| }}{\sigma_n}.
\end{equation}
 Hence the projected noninteracting two-fermion correlation function is given by
\begin{equation}
 C_0^{(rel)} (\tau, \epsilon_n ) =     e^{ -   \frac{k_n^2}{2\mu}  \tau   }      .
\end{equation}

Putting all together, up to leading order, we find
\begin{equation}
 C^{(rel)} (\tau, \epsilon_n ) =    e^{ -   \frac{k_n^2}{2\mu}  \tau   }  \left [1   - \tau  \frac{ \sigma_n V_0}{L}  +   \mathcal{O} ( \frac{V^2_0}{L^2}  ) \right ]      .
\end{equation}
 The leading order contribution of energy level of two-fermion system in a finite volume is therefore obtained by
\begin{equation}
\epsilon_n = - \frac{1}{  \tau} \ln   C^{(rel)} (\tau, \epsilon_n )  =   \frac{k_n^2}{2\mu} +  \frac{\sigma_n V_0}{L} + \mathcal{O} ( \frac{V^2_0}{L^2}  ) .
\end{equation}
The leading effect of energy shift due to interaction in a finite volume in perturbation theory is thus  again given by $ \delta \epsilon_n = \sigma_n \frac{V_0}{L} $.

In terms of perturbation calculation, now we can indeed see   the structure mentioned in Eq.(\ref{tauoverLrelation}),
\begin{equation}
C^{(rel)}(  t  )    -C_0^{(rel)}(  t  )      = - \tau \langle \hat{V} (\tau) \rangle  + \cdots,    
\end{equation}
where 
\begin{equation}
   \langle \hat{V} (\tau) \rangle  =   \sum_{n=0}^\infty  \left (   \frac{ \sigma_n V_0}{L} \right ) e^{ -   \epsilon_n^{(0)}   \tau   }  \propto \frac{1}{L}  .
\end{equation}
Also as demonstrated in  Fig.~\ref{pertbplot}, the difference of integrated correlation functions approaches infinite volume limit much more rapidly in the region of $\tau \ll L$.

\section{Monte Carlo simulation test in 1D quantum mechanics}\label{MCsimulation}  
To demonstrate the feasibility  of proposed formalism,  we  conduct a simple  Monte Carlo simulation test with a 1D   quantum mechanics model in this section.

\subsection{A 1D quantum mechanics model}
 Considering a spinless particle with the mass $\mu$  interacting with a short-range repulsive square well  potential in a harmonic oscillator trap,   the eigen-solutions are determined by  Schr\"odinger equations,
\begin{equation}
\left [ - \frac{1}{2\mu} \frac{d^2}{d r^2} + \frac{1}{2} \mu \omega^2 r^2 +V(r)  \right] \psi_n (r) =\epsilon_n \psi_n (r), 
\end{equation}  
where 
\begin{equation}
V(r) = 
 \begin{cases}  
 \frac{V_0}{b} , &  r \in [-\frac{b}{2}, \frac{b}{2}], \\
 0, & \text{otherwise},
\end{cases}  \stackrel{b \rightarrow 0}{\rightarrow} V_0 \delta(r).
\end{equation}
The energy spectra of the system can be solved by diagonalizing Hamiltonian matrix
\begin{equation}
H_{n, n'} = \delta_{n,n'} \omega (n+\frac{1}{2}) + \frac{V_0}{b}  \int_{-\frac{b}{2}}^{\frac{b}{2}} d r \varphi^{(\omega)*}_{n} (r) \varphi^{(\omega)}_{n'} (r), \label{Hmatsquarewell}
\end{equation}
where $\varphi^{(\omega)}_{n} (r) $ are eigen-solutions of harmonic oscillator potential,
\begin{equation}
\varphi^{(\omega)}_{n} (r) = \frac{1}{\sqrt{2^n n!}} \left ( \frac{\mu \omega}{\pi} \right )^{\frac{1}{4}} e^{- \frac{\mu \omega}{2} r^2} H_n (\sqrt{\mu \omega} r).
\end{equation}

\subsection{Integrated transition amplitude and Monte Carlo simulation}
In Euclidean space-time,  the transition amplitude for a particle propagating from $(r,0)$ to $(r', \tau)$ is defined by
\begin{equation}
\langle r' | e^{- \hat{H} \tau} | r \rangle = \sum_{n=0}^\infty e^{- \epsilon_n \tau}  \psi_n (r) \psi^*_n (r').
\end{equation}
Hence,  integrated transition amplitude (particle propagator)  is associated to the partition function in statistics by,
\begin{equation}
C (\tau) = \int d r \langle r | e^{- \hat{H} \tau} | r \rangle = \sum_{n=0}^\infty  e^{- \epsilon_n \tau} \stackrel{\tau \rightarrow \beta}{ \rightarrow } Tr[e^{ - \beta H}] .
\end{equation}
The path integral representation of the integrated particle propagator is given by, see e.g. Ref.~\cite{Creutz:1980gp,Lepage:1998dt},
\begin{equation}
C (\tau)  = \lim_{N_\tau \rightarrow \infty} \left ( \frac{ \mu}{2\pi a_\tau } \right)^{\frac{N_\tau }{2}} \int \prod_{i=1}^{N_\tau } d r_i e^{- S_E \left (  \{r_i \} \right ) },
\end{equation}
where the time interval $[0, \tau]$ is divided into  $N_\tau$ small steps of width of $a_\tau = \frac{\tau}{N_\tau}$.  The discrete  Euclidean space-time action is given by the sum of  trap action  and interaction term, $S_E \left (\{r_i \} \right) = S^{(\omega)}_E \left (\{r_i \} \right) + S^{(V)}_E \left (\{r_i \} \right)$:
\begin{equation}
S^{(\omega)}_E \left (\{r_i \} \right) = a_\tau \sum_{i =1}^{N_\tau } \left [ \frac{\mu}{2 } \left  ( \frac{r_{i+1} -r_i}{a_\tau} \right )^2  + \frac{1}{2} \mu \omega^2 r_i^2   \right ],  
\end{equation}
and
\begin{equation}
S^{(V)}_E \left (\{r_i \} \right)= a_\tau \sum_{i =1}^{N_\tau}   V(r_i) ,  
\end{equation}
 where $r_0 = r$ and $ r_{N_\tau} = r'  = r$ are initial  and finial position of particle respectively. Similarly when the interaction, $V(r)$, is turned off,  the integrated  particle propagator in the harmonic oscillator trap  is defined by
 \begin{align}
C_0 (\tau)  &= \sum_{n=0}^\infty e^{- \omega (n+ \frac{1}{2}) \tau}  = \frac{1}{2} \mbox{csch} (\frac{\omega \tau }{2}) \nonumber \\
&= \lim_{N_\tau \rightarrow \infty} \left ( \frac{ \mu}{2\pi a_\tau } \right)^{\frac{N_\tau }{2}} \int \prod_{i=1}^{N_\tau } d r_i e^{- S^{(\omega)}_E \left (\{r_i \} \right) }.
\end{align}
  For a finite square well model, without the constraint of Pauli exclusive principle, now both even and odd parity states contribute to $C(\tau)$.

  The path integral representation of ratio of $C(\tau )$ and $C_0 (\tau)$    can be written as
 \begin{equation}
\frac{C (\tau) }{ C_0 (\tau) }  = \lim_{N_\tau \rightarrow \infty}  \int \prod_{i=1}^{N_\tau } d r_i   \rho \left (\{r_i \} \right)  e^{- S^{(V)}_E \left (\{r_i \} \right) } , \label{CoverC0}
\end{equation}
where
\begin{equation}
\rho \left (\{r_i \} \right) = \left ( \frac{ \mu}{2\pi a_\tau } \right)^{\frac{N_\tau }{2}} \frac{ e^{- S^{(\omega)}_E \left (\{r_i \} \right) } }{ C_0 (\tau) }  
\end{equation}
is positive definite and 
\begin{equation}
 \int \prod_{i=1}^{N_\tau} d r_i   \rho \left (\{r_i \} \right)   =1.
\end{equation}
Hence $ \rho \left (\{r_i \} \right) $ can be interpreted as probability density, and Eq.(\ref{CoverC0}) can be computed via standard Monte Carlo simulation method,
 \begin{equation}
\frac{C (\tau) }{ C_0 (\tau) }  = \frac{1}{N_{cfg}}  \sum_{\alpha =1}^{N_{cfg}} e^{- S^{(V)}_E \left (\{r^{(\alpha)}_i \} \right) } , \label{CoverC0MC}
\end{equation}
where $N_{cfg}$ is total number of configurations, $\alpha$ is used to label each configuration,   the random values of $\{r^{(\alpha)}_i \} $ for each individual configuration   can be generated according to the probability density distribution $ \rho \left (\{r^{(\alpha)}_i \} \right) $. The Monte Carlo simulation can be performed rather straightforwardly by standard Metropolis algorithm, see e.g.  Ref.~\cite{Creutz:1980gp,Lepage:1998dt}.

  \begin{figure*}
 \centering
 \begin{subfigure}[b]{0.49\textwidth}
\includegraphics[width=0.99\textwidth]{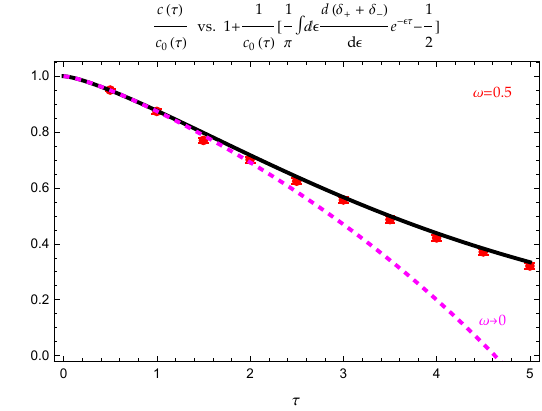}
\caption{  }\label{MCdataplot3}
\end{subfigure} 
\begin{subfigure}[b]{0.49\textwidth}
\includegraphics[width=0.99\textwidth]{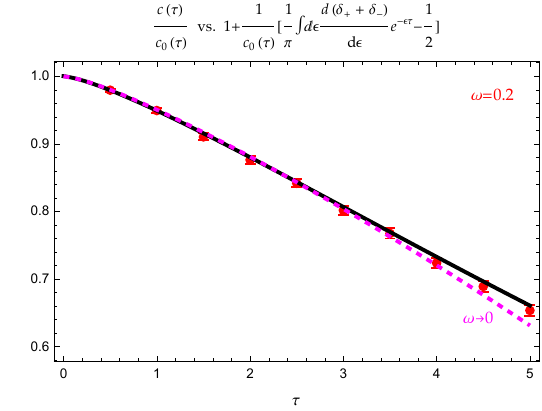}
\caption{    }\label{MCdataplot2}
\end{subfigure}
\begin{subfigure}[b]{0.49\textwidth}
\includegraphics[width=0.99\textwidth]{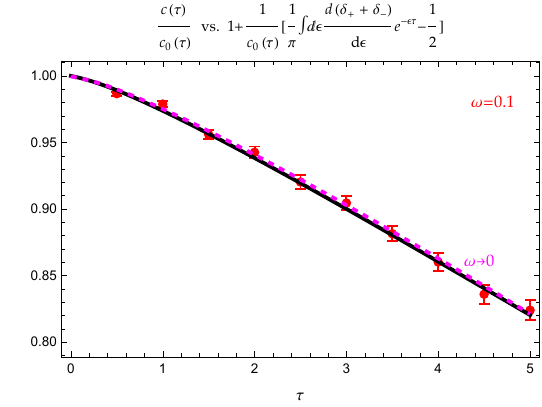}
\caption{    }\label{MCdataplot1}
\end{subfigure}
\begin{subfigure}[b]{0.49\textwidth}
\includegraphics[width=0.99\textwidth]{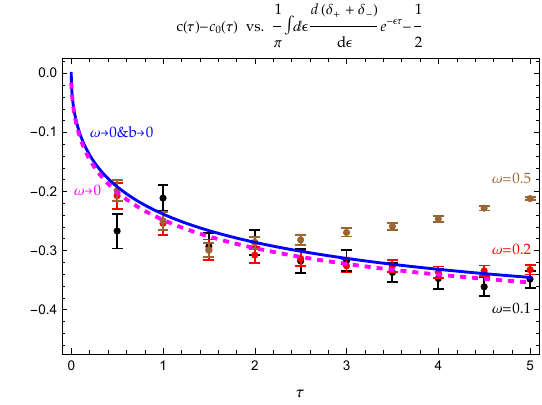}
\caption{    }\label{MCdatawsplot}
\end{subfigure}
 \caption{Comparison of Monte Carlo data of $\frac{C(\tau)}{C_0 (\tau)}$  vs. infinite volume limit result for a spinless particle interacting with a square well potential in a harmonic oscillator trap for various $\omega$'s. In (a), (b) and (c),  we plot Monte Carlo data (red error bars),  exact solutions of   $\frac{C(\tau)}{C_0 (\tau)}$ by solving Hamiltonian matrix in Eq.(\ref{Hmatsquarewell}) (solid black), and  infinite volume limit result, $1 + \frac{1}{C_0 (\tau)} \left [  \frac{1}{\pi}   \int_0^\infty d \epsilon   \frac{  d   \left (\delta_+   + \delta_-   \right ) }{d  \epsilon}   e^{-   \epsilon \tau}   -\frac{1}{2}  \right ]$ (dashed purple), for $\omega =0.5, 0.2$ and $0.1$ respectively.  (d)  Monte Carlo data of $C(\tau) - C_0 (\tau)$ for various $\omega$'s: $\omega =0.1$ (black), $0.2$ (red) and $0.5$ (brown), vs.  infinite volume limit result with a square well potential, $  \frac{1}{\pi}   \int_0^\infty d \epsilon   \frac{  d   \left (\delta_+  + \delta_-    \right ) }{d  \epsilon}   e^{-   \epsilon \tau}   -\frac{1}{2} $ (dashed purple). The infinite volume limit result with a contact interaction potential, $V(r) = V_0 \delta(r)$, is also plotted as a comparison in solid blue.    The rest of parameters are taken as:       $V_0=1$, $\mu=1$ and $b=0.2$.  \label{MCdatafigs} }
 \end{figure*}

\subsection{Scattering phase shifts and its relation to integrated transition amplitude}
The scattering amplitudes for a repulsive square well potential  in infinite volume can be solved analytically, see e.g. Appendix D in Ref.~\cite{Guo:2016fgl}. The phase shifts, $\delta_{\pm} (\epsilon)$, are given by
\begin{equation}
   \delta_{\pm} (\epsilon)  =  \cot^{-1} \left [ \frac{  1+ \left ( \frac{k}{k_V} \right )^{\pm 1}  \cot \left  (\frac{k b}{2} \right  )    \cot \left  (  \frac{ k_V b}{2} \right )    }{\cot \left  (\frac{k b}{2} \right )  -  \left ( \frac{k}{k_V} \right )^{\pm 1}   \cot \left  (  \frac{ k_V b}{2} \right )    }   \right ], \label{phaseshiftsquarewell}
\end{equation}
where subscripts $(+/-)$ are used to label even and odd parity states respectively,  and $$k =\sqrt{2\mu \epsilon}, \ \ \ \   k_V = \sqrt{2\mu \left  (\epsilon- \frac{V_0}{b} \right )}.$$ As $b \rightarrow 0$, 
\begin{equation}
 \delta_{+} (\epsilon)  \stackrel{ b\rightarrow 0}{\rightarrow} \cot^{-1} \left ( - \frac{k}{\mu V_0} \right ), \ \ \ \  \delta_{-} (\epsilon)  \stackrel{ b\rightarrow 0}{\rightarrow}  0,
\end{equation}
the square well potential approaches a contact interaction, and  solutions for odd parity states are suppressed. The scattering phase shifts are associated to $C (\tau) $ and $ C_0 (\tau)$ by
\begin{align}
& C (\tau)  - C_0 (\tau) \nonumber \\
& \stackrel{\omega \rightarrow 0}{\rightarrow} \frac{1}{\pi}   \int_0^\infty d \epsilon  \left [  \frac{  d  \delta_+ (\epsilon)  }{d  \epsilon}  + \frac{  d    \delta_- (\epsilon) }{d  \epsilon}  \right ] e^{-   \epsilon \tau}  +  \frac{\delta_+(0)+\delta_-(0) }{\pi} , \label{CthotrapMC}
\end{align}
where $\delta_+(0) = - \frac{\pi}{2}$ and $\delta_-(0)=0$.

\subsection{Monte Carlo data vs. exact solutions vs. infinite volume limit result}
Numerical test for the system of a spinless particle interacting with a square well potential  are carried out and presented in this subsection,  the aim is to demonstrate that Monte Carlo result of trapped system  approaches and converge with infinite volume limit result   at small Euclidean time region.

(i) The Monte Carlo computation of $\frac{C (\tau) }{ C_0 (\tau) }   $ in Eq.(\ref{CoverC0MC}) for the system in a harmonic oscillator trap is carried out by standard Metropolis algorithm, see e.g.  Ref.~\cite{Creutz:1980gp,Lepage:1998dt}. The simulation are performed with fixed number of steps in temporal dimension, $N_\tau=100$, so the  lattice spacing $a_\tau = \frac{\tau}{N_\tau}$ varies for   $\tau \in  [0.5, 5]$. The typical half million measurements are generated for each $\tau$.  The  choice of other parameters are  $V_0 =1$, $\mu =1$ and $b=0.2$ for a square well potential, and various $\omega$'s  for a harmonic oscillator trap are used in our simulation: $\omega  = 0.1,0.2$ and $0.5$. The variance of data samples are computed by Jackknife resampling method.

(ii) As a comparison, the energy spectra of particle interacting with a square well potential in a harmonic oscillator trap can be solved by diagonalizing Hamiltonian matrix in Eq.(\ref{Hmatsquarewell}), so the exact solution of  $\frac{C (\tau) }{ C_0 (\tau) }   $  can be obtained, where $C(\tau ) = \sum_{n=0}^\infty e^{- \epsilon_n \tau}$ and $C_0 (\tau) = \frac{1}{2} \mbox{csch} (\frac{\omega \tau}{2})$.

(iii) The scattering of a spinless particle off a square well potential in infinite volume can be solved exactly, the analytic expression of phase shifts for both parity states are given in Eq.(\ref{phaseshiftsquarewell}). At infinite volume limit,  
\begin{equation}
 \frac{C (\tau) }{ C_0 (\tau) }   \stackrel{\omega \rightarrow 0}{\rightarrow } 1+  \left [  \frac{1}{\pi}   \int_0^\infty d \epsilon    \frac{  d  \left (  \delta_+  +  \delta_-    \right )  }{d  \epsilon}     e^{-   \epsilon \tau}   - \frac{1}{2} \right ] \omega \tau  .
 \end{equation}
For a small $b \sim 0.1$, the scattering solutions of a square well potential agree well with scattering solution of a contact interaction.

The comparison of Monte Carlo data of $ \frac{C (\tau) }{ C_0 (\tau) } $ (red error bars) vs. exact solutions (solid black) vs. infinite volume limit result (dashed purple) are shown in Figs.~\ref{MCdataplot3}-\ref{MCdataplot1} for various $\omega$'s.  The Monte Carlo data of $ C (\tau) - C_0 (\tau)  $ for various $\omega$'s  (colored error bars)  vs.  infinite volume limit result (dashed purple) are shown in  Fig.~\ref{MCdatawsplot}, the infinite volume limit result with a contact interaction (solid blue) is also plotted in Fig.~\ref{MCdatawsplot} as a comparison.

\section{Discussion and summary}\label{summary}

In summary, a relation between  integrated correlation function of a trapped system and infinite volume  scattering phase shift is derived in present work. We show that even with a modest size of a trap,  the difference of integrated correlation function of a trapped system with and without particles interactions  at small Euclidean time region approach steadily  to its infinite volume limit that is given in terms of scattering phase shift by
\begin{equation}
  C  (t)  - C_0 (t)     \stackrel[t=-i\tau]{trap \rightarrow \infty}{\rightarrow} \frac{1}{\pi}   \int_0^\infty d \epsilon  \frac{  d \delta (\epsilon) }{d  \epsilon}   e^{-    \epsilon  \tau}  +  \frac{ \delta(0)}{\pi}.    \nonumber
\end{equation}
Therefore, the scattering phase shifts may be extracted from lattice  simulation of integrated correlation function at small time region, which is in great contrast to conventional two-step approach in extracting scattering informations from  lattice calculation:   extracting energy levels from temporal correlation function in large Euclidean time region in the first step and  then converting energy spectra into phase shifts by applying L\"uscher formula in the second step.    Both (1)  perturbation calculation of  1+1D lattice Euclidean field theory model of fermions interacting with a contact  interaction and  (2) Monte Carlo simulation  of a 1D exactly solvable quantum mechanics model are carried out to explore and test the proposed relation,   we show both analytically and numerically  that  the difference of integrated correlation function of a trapped system  indeed agree well with infinite volume limit at small time region even for a modest small size of trap.

The fundamental reason of this observation is due to the fact that   integrated  trapped correlation functions   resemble the partition function  in statistical mechanics, 
\begin{equation}
  C  (t)  - C_0 (t)  \stackrel{t = - i \tau}{  \rightarrow}  Tr[e^{ -   \hat{H}  \tau} -e^{ -   \hat{H}_0 \tau  }  ],   \nonumber 
  \end{equation}
with $\tau$  playing the role of the square of the thermal de Broglie wavelength. When thermal de Broglie wavelength is much smaller than size of trap, particles are nearly blind of   the  size effect of a trap,   the difference of integrated trapped two-particle correlation functions  can be described in terms of power of $\tau/L$,  the leading order contribution is thus proportional to $\tau  \langle \hat{V} \rangle$ where $   \langle \hat{V} \rangle \sim 1/L$.

 The scope of  current discussion is still limited to 1+1D nonrelativistic few-particle dynamics.  The current focus of  this work is to simply   demonstrate  both numerically and analytically that   the difference of integrated trapped two-particle correlation functions converges quickly to its infinite volume limit which is expressed through an integral over the derivative of the phase shift weighted by an exponential factor. The  fast convergent  feature of proposed relation near small Euclidean times may have the potential to provide an alternative approach to traditional two-step L\"uscher formula method. The ultimate goal is to develop an alternative method that can be a robust tool to extract phase shift from LQCD calculation especially in  cases when the traditional two-step L\"uscher formula method becomes less effective and determination of individual energy levels itself is already problematic, such as in nucleon-nucleon reactions. Much further work is required to accomplish  this ultimate goal.    The proposed approach will have to be extended to include relativistic dynamics, inelastic effect, etc.  Monte Carlo simulation  with field theory models are also demanded for the effectiveness and robustness test. We  also  remark that unlike L\"uscher formula approach that relate energy levels to phase shift directly, our proposed approach requires the physics motivated modeling of phase shift and then fit to the LQCD data to fix model parameters. This resembles the procedure of determination of   hadron-hadron scattering amplitudes  from experimental data, see e.g. Ref.\cite{Ananthanarayan:2000ht}. The model of phase shift in principle can be further constrained by chiral perturbation theory, dispersion relation approach, Roy equation, etc., which can ultimately help to narrow down the parameter space in the model.

\acknowledgments
We   acknowledges support from  the College of Arts and Sciences,  Dakota State University, Madison, SD  and  the Department of Physics and Engineering, California State University, Bakersfield, CA.
V.G. thanks UPCT for partial financial support through the concession of "Maria Zambrano ayudas para la recualificación del sistema universitario español 2021-2023" financed by Spanish Ministry of Universities with financial funds "Next Generation" of the EU.  This research was supported in part by the National Science Foundation under Grant No. NSF PHY-1748958.

\appendix

\section{Quantization condition of trapped two-particle system in 1+1D}\label{QC1Dappend}
Some technical details of derivation of quantization condition of trapped two-particle system in 1+1D are given in this section. Let's consider two particles interact with a contact interaction in a trap, two particles could be either spinless particles or two fermions in spin singlet state. Hence only even parity state will  be affected by contact interaction. The dynamics of relative motion of trapped two-particle systems are described by: (1)  Eq.(\ref{schrodingerbox})  for a periodic box,  wave function satisfies periodic boundary condition of wave function   in Eq.(\ref{periodicwave});  and (2)  Eq.(\ref{schrodingerhotrap}) for a harmonic oscillator trap.

The integral representation of dynamics of trapped systems are given by Lippmann-Schwinger equation,
\begin{equation}
\psi_\epsilon^{(rel)} (r) = \int_{trap} d r' G^{(trap)}_0 (r,r'; \epsilon) V_0 \delta (r') \psi_\epsilon^{(rel)} (r') ,
\end{equation} 
where
\begin{equation}
\int_{trap} d r'  = \begin{cases} \int_{-\frac{L}{2}}^{\frac{L}{2}} d r' ,  & \mbox{for a periodic box},  \\ \int_{- \infty}^{ \infty} d r' ,  & \mbox{for a h.o. trap}.   \end{cases}
\end{equation}
The $G^{(trap)}_0 (r,r'; \epsilon) $ is Green's function of non-interacting particles in a trap:  (1) for a periodic box, it satisfies differential equation,
\begin{equation}
 \left [ \epsilon + \frac{1}{2\mu } \frac{d^2}{d r^2}  \right ] G_0^{(trap)}(r,r'; \epsilon)   = \sum_{n \in \mathbb{Z}} \delta(r-r' + n L),
\end{equation}
and periodic boundary condition, $$G_0^{(trap)}(r + n L,r'; \epsilon) = G_0^{(trap)}(r,r'; \epsilon);$$ and  (2) for a harmonic oscillator trap, it satisfies equation,
\begin{equation}
 \left [ \epsilon + \frac{1}{2\mu } \frac{d^2}{d r^2} - \frac{1}{2} \mu \omega^2 r^2 \right ] G_0^{(trap)}(r,r'; \epsilon)   = \delta(r-r').
\end{equation}
Hence the discrete energy spectra are determined by quantization condition,
\begin{equation}
\frac{1}{V_0}=  G^{(trap)}_0 (0,0; \epsilon)  .
\end{equation}

The analytic expression of $G_0^{(trap)}(r,r'; \epsilon)$ can be obtained for both periodic box and harmonic oscillator trap:

 (1) for a periodic box, see e.g. Refs.~\cite{Guo:2013vsa,Guo:2016fgl,Guo:2019hih}
\begin{align}
& G_0^{(trap)}(r,r'; \epsilon)  = \frac{1}{L} \sum^{p = \frac{2\pi n}{L}}_{n \in \mathbb{Z}} \frac{e^{i p (r-r')}}{ \epsilon - \frac{p^2}{2\mu}} \nonumber \\
&= - \frac{ i \mu}{k} \left [e^{ i k |r-r'|} + \frac{2 \cos k (r-r')}{e^{- i k L}-1} \right ], \label{G0Lperiodicbox}
\end{align}
where $k = \sqrt{2\mu \epsilon}$; 

(2) for a harmonic oscillator trap, see e.g. Ref.~\cite{Blinder83},
\begin{align}
&G_0^{(trap)}(r,r'; \epsilon)  \nonumber \\
&= - \frac{\Gamma(\frac{1}{4} - \frac{\epsilon}{2\omega} )}{2\omega  (\pi r r')^{\frac{1}{2}}}  M_{\frac{\epsilon}{2\omega}, -\frac{1}{4}} (\mu \omega r^2_{<}) W_{\frac{\epsilon}{2\omega},-\frac{1}{4}} (\mu \omega r^2_{>}),\label{W}
\end{align}
where $M$ and $W$ are Whittaker functions as defined in  Ref.~\cite{whittaker_watson_1996}, and $r_{<}$ and $r_>$ represent the lesser and greater of $(r,r')$ respectively.

 We thus find
\begin{equation}
G_0^{(trap)}(0,0; \epsilon) = \begin{cases}  \frac{  \mu}{k} \cot (\frac{k L}{2}), & \mbox{for a periodic box}, \\
 - \frac{\sqrt{\mu}}{2 \sqrt{\omega} }  \frac{\Gamma(\frac{1}{4} - \frac{\epsilon}{2\omega} )}{ \Gamma(\frac{3}{4} - \frac{\epsilon}{2\omega} )}, & \mbox{for a h.o. trap}.\end{cases}
\end{equation}
The analytic expression of scattering phase shift for a contact interaction is given by Eq.(\ref{deltaphase}),
$$ \frac{1}{  V_0}    =   -  \frac{\mu  }{k} \cot \delta(\epsilon). $$ Hence the quantization conditions for (1) a periodic box and (2) a harmonic oscillator trap can be written in the form of  L\"uscher formula  \cite{Luscher:1990ux}    and BERW formula \cite{Busch98}, 
\begin{equation}
  \cot \delta(\epsilon)  - \mathcal{M} (\epsilon) = 0,  
\end{equation}
where
\begin{equation}
  \mathcal{M} (\epsilon) =  \begin{cases}  - \cot ( \sqrt{2\mu \epsilon} \frac{ L}{2}), & \mbox{for a periodic box}, \\
\frac{\sqrt{\mu}}{2 \sqrt{\omega} }  \frac{\Gamma(\frac{1}{4} - \frac{\epsilon}{2\omega} )}{ \Gamma(\frac{3}{4} - \frac{\epsilon}{2\omega} )}, & \mbox{for a h.o. trap}.\end{cases}
\end{equation}

It is worth mentioning that the superlattice  structure that has great interest in condensed matter physics can be constructed by placing finite number of short-range interaction potentials inside of traps. For an example, assuming potential of a simple superlattice structure having the form of
\begin{equation}
V(r) = \sum_{i = 1}^N V_{i} \delta (r - a_i),
\end{equation}
where $a_i  $ is location of $i$-th contact potential in the trap, the quantization condition that determine discrete energy spectra can be   obtained from Lippmann-Schwinger equation:
\begin{equation}
\det \left [ \delta_{i,j}  - V_j G_0^{(trap)} (a_i, a_j; \epsilon) \right ] =0. \label{superlatticeQC}
\end{equation}
With some mathematical manipulation, we can easily show that the QC in Eq.(\ref{superlatticeQC}) is consistent with the result that is derived from   the characteristic determinant approach  in Refs. \cite{vg1,vg2}: $$\det[D_N]=0,$$ where 
\begin{equation}
(D_N)_{i,j}=\delta_{i,j}-V_j G_0^{(trap)}(a_j,a_j; \epsilon) \sqrt{Z_{i,j}} ,
\end{equation}
and
\begin{equation}
  Z_{i,j} =Z_{j,i}=     \frac{G_0^{(trap)}(r_i,r_j; \epsilon)G_0^{(trap)}(r_j,r_i; \epsilon)}{G_0^{(trap)}(r_i,r_i; \epsilon)G_0^{(trap)}(r_j,r_j; \epsilon)}.
\end{equation}
We remark that the determinant in Eq.(\ref{superlatticeQC}) is directly related to the full Green’s function of the system.  The  full Green’s function of the system  provides the transmission coefficient and the density of states when opened
systems are considered, and gives the bound spectrum if
the system is closed (see Refs. \cite{vg1,vg2} for more details).  The poles of Green’s function   are the zeros of the determinant in Eq.(\ref{superlatticeQC}) for a closed system.
The   characteristic determinant approach is a convenient formalism to
determine the energy spectrum electrons in a layered system or get sufficiently
complete description of electron behavior in a random potential without finding electron eigenfunctions.

\section{Friedel formula and Krein's theorem in 1+1D scattering theory}\label{FriedelKreinappend}
A brief review of Friedel formula and Krein's theorem in 1+1D scattering theory is provided in this section, detailed discussion can be found in Refs.~\cite{doi:10.1080/00018735400101233,Friedel1958,krein1953trace,zbMATH03313022,Faulkner_1977}. The derivation can also be made in a rather more general  way  from formal scattering theory and $S$-matrix  formulation approach, see e.g. Refs.~\cite{PhysRevA.6.851.2,Guo:2022row}.

In Refs.~\cite{doi:10.1080/00018735400101233,Friedel1958}, J. Friedel showed that the difference between the integrated density of states of the interacting particles system, $n_E (x) $, and free particles system, $n_E^{(0)} (x)$, is related to the scattering phaseshifts by
\begin{equation}
\int_{-\infty}^\infty d x \left [n_E (x) - n_E^{(0)} (x) \right ] = \frac{1}{ \pi} \frac{d }{d E} Tr \left [ \delta (E) \right ], \label{Friedelformula}
\end{equation}
where $\delta(E)$ stands for the diagonal matrix of scattering phaseshifts.  The local density of states of a interacting system, $n_E(x)$, can be defined through the imaginary part of Green's function by
\begin{equation}
n_E (x) = - \frac{1}{\pi} Im  \left [ \langle x | \hat{G}(E+ i 0 )| x \rangle \right ],
\end{equation}
where $$\hat{G}(E) = \frac{1}{E- \hat{H}}$$ refers to full  Green's function operator of an interacting particles system, and   $\hat{H}  $ stands for the Hamiltonian operator of the interacting particles  system.  The local density of states of free particles system, $n_E^{(0)} (x)$, is defined in a similar way,
\begin{equation}
n^{(0)}_E (x) = - \frac{1}{\pi} Im  \left [ \langle x | \hat{G}_0(E+ i 0 )| x \rangle \right ],
\end{equation}
where $$\hat{G}_0(E) = \frac{1}{E- \hat{H}_0}$$ denotes the    free particle's Green's function operator. 
The relation   in 
Eq.(\ref{Friedelformula})   is usually referred as the Friedel formula.

   The real part (principal part) of Green's function   can be constructed through imaginary part by Cauchy's integral theorem,   
\begin{align}
&   \int_{-\infty}^\infty d x   \langle x | \hat{G}(E )  -  \hat{G}_0(E )| x \rangle    \nonumber \\
& =    \frac{1}{\pi} \left [ \int_{-\infty}^{-E_L} +\int_{0}^\infty \right ] d \epsilon \frac{  \int_{-\infty}^\infty d x  Im  \langle x | \hat{G}(E )  -  \hat{G}_0(E )| x \rangle   }{\epsilon - E}   , 
\end{align}
where we  have assumed that Green's functions    has a physical branch cut along the positive real axis in complex $E$-plane: $E \in [0, \infty]$,   and an unphysical branch cut siting along negative real axis: $E \in [- \infty, -E_L]$,   where $-E_L$ represents the branch point of unphysical cut.   Using Eq.(\ref{Friedelformula}), we   therefore find that  integrated Green's function is related to the scattering phaseshifts by
\begin{align}
 &  \int_{-\infty}^\infty d x   \langle x | \hat{G}(E )  -  \hat{G}_0(E )| x \rangle   \nonumber \\
 &  =  -   \frac{1}{\pi} \left [ \int_{-\infty}^{-E_L} +\int_{0}^\infty \right ] d \epsilon \frac{  Tr [\delta(\epsilon)]   }{(\epsilon - E)^2}   . \label{Kreinformula}
\end{align}
J.S. Faulkner \cite{Faulkner_1977} later on recognized that  the relation   in Eq.(\ref{Kreinformula}) is equivalent to Krein's theorem \cite{krein1953trace,zbMATH03313022} in spectral theory, where 
$ - \frac{1}{ \pi }  Tr [\delta(\epsilon)]  $ is exactly the  Krein's spectral shift function.

\bibliography{ALL-REF.bib}

\end{document}